\pdfoutput=1

\documentclass[submission, Phys]{SciPost}

\usepackage[utf8]{inputenc}
\usepackage[T1]{fontenc}
\usepackage{amsmath, amsfonts, amssymb}
\usepackage{mathtools}
\usepackage{physics}
\usepackage{graphicx}
\usepackage{placeins}
\usepackage{svg}
\usepackage{subcaption}
\usepackage{slashed}
\usepackage{multirow}

\usepackage{booktabs}
\usepackage{array}
\usepackage{makecell}

\DeclareUnicodeCharacter{27E8}{\langle}
\DeclareUnicodeCharacter{27E9}{\rangle}
\DeclareUnicodeCharacter{00B2}{^2}

\usepackage[ruled,lined]{algorithm2e}  
\SetKwComment{Comment}{/* }{ */}

\newcommand{\ii}{\,\mathrm{i}\,}

\graphicspath{{figures/}}

\usepackage{easy-todo}
\usepackage{cleveref}
\usepackage{color}
\usepackage{bbm}
\usepackage{enumitem}

\def\NC{{N_c}}
\def\NF{{N_f}}
\def\NFqq{{\tilde{N}_f}}
\DeclareMathOperator{\Span}{span}

\newcommand{\MSbar}{\overline{\text{MS}}}

\def\trFive{{\rm tr}_5}

\usepackage[bitstream-charter]{mathdesign}
\DeclareSymbolFont{usualmathcal}{OMS}{cmsy}{m}{n}
\DeclareSymbolFontAlphabet{\mathcal}{usualmathcal}

\begin{document}

\begin{flushright}
  \footnotesize
  CERN-TH-2023-090, 
  PSI-PR-23-16,
  ZU-TH 23/23
\end{flushright}

\begin{center}{\Large \textbf{
Two-Loop QCD Corrections for Three-Photon Production at Hadron Colliders\\
}}\end{center}

\begin{center}
Samuel~Abreu\textsuperscript{1,2},
Giuseppe De Laurentis\textsuperscript{3},
Harald Ita\textsuperscript{3,4},
Maximillian~Klinkert\textsuperscript{5},
Ben Page\textsuperscript{1} and
Vasily Sotnikov\textsuperscript{6,7}
\end{center}

\begin{center}
{\bf 1} CERN, Theoretical Physics Department, CH-1211 Geneva 23, Switzerland
\\
{\bf 2} 
Higgs Centre for Theoretical Physics, School of Physics and Astronomy,\\
The University of Edinburgh, Edinburgh EH9 3FD, Scotland, UK
\\
{\bf 3}
Paul Scherrer Institut, CH-5232 Villigen PSI, Switzerland
\\
{\bf 4}
ICS, University of Zurich, 
Winterthurerstrasse 190, 8057 Zurich, Switzerland
\\
{\bf 5} 
Physikalisches Institut, Albert-Ludwigs-Universit\"at Freiburg,\\
Hermann-Herder.~Str.~3, D-79104 Freiburg, Germany
\\
{\bf 6} 
Physik-Institut, University of Zurich, \\
Winterthurerstrasse 190, 8057 Zurich, Switzerland
\\
{\bf 7} 
Department of Physics and Astronomy, Michigan State University,\\
567 Wilson Road, East Lansing, MI 48824, USA
\\
\end{center}

\begin{center}
\today
\end{center}


\section*{Abstract}
{
We complete the computation of the two-loop helicity amplitudes for the
production of three photons at hadron colliders, 
including all contributions beyond the leading-color approximation.
We reconstruct the analytic form of the amplitudes from numerical finite-field 
samples obtained with the numerical unitarity method. 
This method requires as input surface terms for all relevant five-point non-planar integral 
topologies, which we obtain by solving the associated syzygy problem in embedding space.
The numerical samples are used to constrain compact spinor-helicity ansätze, 
which are optimized by taking advantage of the known one-loop analytic 
structure.
We make our analytic results available in a public \texttt{C++} library, 
which is suitable for immediate phenomenological applications.
We estimate that the inclusion of the subleading-color contributions will 
decrease the size of the two-loop corrections by about $30\%$ to $50\%$,
and the NNLO cross sections by a few percent,
compared to the results in the leading-color approximation.
}

\newpage
\vspace{10pt}
\noindent\rule{\textwidth}{1pt}
\tableofcontents\thispagestyle{fancy}
\noindent\rule{\textwidth}{1pt}
\vspace{10pt}



\section{Introduction}
\label{sec:intro}

With the rapid development of precision studies within the physics
program at the Large Hadron Collider (LHC), there is a growing
need for precise theory predictions for many Standard Model processes.
The knowledge of higher-order radiative corrections in the strong coupling
constant is essential for the correct interpretation of the associated 
experimental measurements, and provides better control over 
theoretical uncertainties. 
While many two-to-two reactions are now known at next-to-next-to-leading order 
(NNLO) in perturbation theory, NNLO theoretical predictions 
for two-to-three processes constitute the current state of the art.

The study of triphoton final states at hadron colliders
such as the LHC offers excellent opportunities for several interesting
investigations. For instance, it allows to explore the consequences of 
potentially anomalous gauge and Higgs couplings \cite{ATLAS:2015rsn,
Denizli:2019ijf,Denizli:2020wvn,Aguilar-Saavedra:2020rgo,Senol:2023fpa}.
Furthermore, the study of triphoton production is an important irreducible 
background process for the associated production of a photon 
with a beyond-the-Standard-Model particle that subsequently decays 
into a pair of photons \cite{Zerwekh:2002ex,Toro:2012sv,Das:2015bda}.
Similar to diphoton production \cite{Catani:2011qz,Campbell:2016yrh,Catani:2018krb},
triphoton production exhibits large next-to-leading order (NLO) and NNLO
corrections \cite{Chawdhry:2019bji,Kallweit:2020gcp,ATLAS:2021lsc}.
Accordingly, NLO predictions significantly deviate from data 
\cite{ATLAS:2017lpx}, and NNLO is the first perturbative order providing 
reliable results \cite{Kallweit:2020gcp,Chawdhry:2019bji} 
(see also the related work of refs.~\cite{Valeshabadi:2021twn,
Karpishkov:2022ukm}).
Thus, good control of NNLO QCD corrections to this process is of great
importance. 

While diphoton production has been known at NNLO QCD accuracy for more than 
a decade~\cite{Catani:2011qz,Campbell:2016yrh}, first results for
triphoton production at the same level of accuracy have only 
recently been obtained~\cite{Chawdhry:2019bji,Kallweit:2020gcp}.
This is due to the highly challenging nature of NNLO computations for 
five-particle scattering processes.
It has been demonstrated that existing frameworks for the subtraction of 
infrared divergences at NNLO are, in principle, capable of 
handling arbitrary production process 
\cite{Kallweit:2020gcp,Czakon:2021mjy,Chen:2022ktf,Catani:2022mfv}. However,
the calculation of two-loop amplitudes for five-particle processes represents 
the current state of the art, and needs to be addressed on a case-by-case 
basis. While NNLO QCD corrections for most massless five-particle processes 
have already been considered \cite{Czakon:2021mjy,Chen:2022ktf,
Chawdhry:2021hkp,Chawdhry:2019bji,Kallweit:2020gcp,Badger:2021ohm,
Badger:2021imn,Badger:2023mgf}, in most of these studies, 
including the NNLO QCD corrections to triphoton production 
\cite{Kallweit:2020gcp,Chawdhry:2019bji},
only the leading-color approximation of the double-virtual corrections 
has been employed.
The first complete cross-section calculation that does not employ the leading-color 
approximation was performed in ref.~\cite{Badger:2023mgf},
and a number of complete two-loop five-point amplitudes are also available 
\cite{Agarwal:2021vdh, Badger:2021imn, Badger:2023mgf}.

Subleading-color corrections are more challenging to compute because 
they include non-planar Feynman graphs, which are notoriously more 
challenging to handle.
In pure QCD, all non-planar contributions vanish in the limit of a large number of colors $\NC$ \cite{tHooft:1973alw}.
This still holds if the number of fermions flavors $\NF$ is considered to 
be of the order of $\NC$, i.e.~when the ratio $\NF/\NC$ is kept constant as 
$\NC$ approaches infinity.
The latter variant of the leading-color approximation is typically 
phenomenologically justified, provided errors of about 10\% in the 
approximated contributions are deemed acceptable. On the other hand,
terms originating from photons coupling to closed fermion loops involve 
nonplanar diagrams. While these terms are still suppressed in the 
\textit{formal} $\NC \to \infty$ limit, 
one might be concerned that their contribution is not suppressed numerically.

The goal of this work is to compute analytic expressions for all two-loop 
five-point amplitudes that contribute to
the NNLO corrections for triphoton production at hadron colliders in
massless QCD (ignoring top loops), and provide an efficient numerical implementation for use in
phenomenological studies.
Our calculation follows the multi-loop numerical unitarity
approach~\cite{Ita:2015tya,Abreu:2017xsl,Abreu:2017hqn,Abreu:2018zmy} as implemented in \textsc{Caravel} \cite{Abreu:2020xvt}.
In numerical unitarity, amplitudes are reduced to a set of master integrals
by matching numerical evaluations of 
generalized unitarity cuts to a parametrization of the loop integrands.
Analytic expressions can then be reconstructed using
multivariate functional reconstruction 
techniques~\cite{vonManteuffel:2014ixa,Peraro:2016wsq} (see also 
refs.~\cite{Klappert:2019emp, DeLaurentis:2019bjh,DeLaurentis:2022otd,
Magerya:2022hvj,Belitsky:2023qho} for related developments).
In order to perform our calculation within this framework, we make a number of
theoretical developments.

Firstly, we develop a new approach to the problem of parametrising the integrand so that as many
terms as possible vanish upon integration, due to integration-by-parts (IBP)
identities \cite{Tkachov:1981wb,Chetyrkin:1981qh}.
These terms are commonly referred to as \emph{surface terms},
and it is a highly
non-trivial problem to construct them in a form that enables efficient
numerical evaluations.
In this work, we present a novel method of deriving numerically efficient
representations of surface terms.
We achieve this by solving an associated syzygy problem
\cite{Gluza:2010ws,Ita:2015tya,Larsen:2015ped} which, taking inspiration from
refs.~\cite{Caron-Huot:2014lda,Bern:2017gdk,Abreu:2017ptx}, is formulated
in so-called embedding space.
This allows us to identify a simpler \emph{homogeneous} syzygy system, 
whose solutions we lift to full syzygies via linear algebra.
By performing this calculation on a numerical phase-space point, we are able to
construct ``skeleton'' syzygies, that allow us to numerically 
determine the full set of syzygies and surface terms
phase-space point by phase-space point in an efficient manner.
This significantly reduces the expression size of surface terms, allowing us to
efficiently match the integrand parametrisation to generalized unitarity cuts.

Secondly, we tackle the important problem of the large amount of samples
required to perform analytic reconstruction.
Indeed, while several improvements to the generic black-box
reconstruction~\cite{vonManteuffel:2014ixa,Peraro:2016wsq} have been
explored~\cite{Abreu:2018jgq,Badger:2021imn,Abreu:2021asb} over the years, 
cutting-edge calculations (see e.g.~\cite{Abreu:2021asb}) indicate that further
developments will be important to tackle amplitudes with an increased number 
of scales.
For this reason, we explore new techniques to construct ansätze whose analytic
structure better exhibit the physical properties of the scattering amplitudes,
performing the analytic reconstruction using spinor-helicity ansatz
techniques~\cite{DeLaurentis:2019bjh,DeLaurentis:2022otd}.
In contrast to a more traditional reconstruction using a set of independent
kinematic invariants (as e.g.\ in \cite{Abreu:2020cwb}), this better manifests
physical properties of the amplitudes. Combined with a further optimization of
the ansatz based on the expectation that some features of the analytic 
structure of one-loop amplitudes
are preserved at two loops, we find a
significant reduction in the number of numerical samples that is required. 
In practice, we are able to
reconstruct the most complicated helicity amplitude from only about 4000
evaluations, corresponding to an order of magnitude less than what was originally 
required for the reconstruction of the planar amplitudes in 
ref.~\cite{Abreu:2020cwb}.

Alongside this paper, we provide the analytic results for the complete two-loop
triphoton production amplitudes in a collection of supplementary material.
Furthermore, in order to facilitate the applicability of our results in
phenomenological studies of triphoton production, we have implemented them in
the efficient public C++ library
\texttt{FivePointAmplitudes}~\cite{FivePointAmplitudes}. This further allows us
to analyze the important question of the impact of the subleading-color
contributions on the two-loop corrections.
Our study suggests that including these contributions will lead to a 
significant decrease in the size of the two-loop corrections, reducing 
them by approximately 30\% to 50\% compared to the results obtained in 
the leading-color approximation. This effect is larger than the corrections 
of about 10\% expected from typical color-suppressed contributions. This 
confirms the concerns that nonplanar contributions arising from the 
photons coupling to closed fermion loops are not necessarily 
numerically suppressed. We note nevertheless that this substantial
change in the two-loop corrections should be contrasted against the 
fact that the double-virtual contributions to the NNLO corrections to this
process are observed to be small \cite{Chawdhry:2019bji,Kallweit:2020gcp}.

The paper is organized as follows. In \cref{sec:notation} we classify the full
set of gauge-invariant contributions to triphoton production.
In \cref{sec:calculation} we discuss our computational approach. We review
numerical unitarity, discuss our approach to the construction of surface terms
and how we construct compact spinor-helicity ans\"atze.
In \cref{sec:results}, we discuss the structure of our results, their 
validation, and the format in which they are presented in ancillary files. 
We also showcase the numerical performance of our \texttt{C++} implementation. In \cref{sec:H-functions-effect}, we discuss the impact of the
subleading-color contributions on the double-virtual corrections.
Finally, in \cref{sec:conclusion}, we present our conclusions.


\section{Notation and Conventions}
\label{sec:notation}

We consider the $\order{\alpha_s^2}$ corrections to the production of three
photons at hadron colliders. The loop-induced process
$gg\to \gamma\gamma\gamma$ vanishes to all orders 
in the combined theory of QCD and QED due to charge-conjugation symmetry \cite{Furry:1938zz}.  
Therefore, the only contributing partonic process is
\begin{equation}\label{eq:process}
  q(-p_1,-h_1)+\bar{q}(-p_2,-h_2)\,\to\,\gamma(p_3,h_3)+\gamma(p_4,h_4)+\gamma(p_5,h_5)\,,
\end{equation} 
where $p_i$ and $h_i$ denote the momentum and the helicity of
the $i^{\text{th}}$ particle, respectively. Throughout this paper,
momenta and helicity labels are understood in the all-outgoing convention.

The process involves five massless particles. Thus, the underlying
kinematic is specified by five Mandelstam invariants, which can be
chosen to be
\begin{align}\begin{split}\label{eq:mand}
  s_{12}=(p_1+&\,p_2)^2\,,\quad s_{23}=(p_2+p_3)^2\,,\quad s_{34}=(p_3+p_4)^2\,,\\
  &s_{45}=(p_4+p_5)^2\,,\quad s_{15}=(p_1+p_5)^2\,,
\end{split}\end{align}
as well as the parity-odd contraction of four momenta,
\begin{equation}\label{eq:tr5}
  \trFive = \trace(\gamma^5\slashed{p}_1\slashed{p}_2\slashed{p}_3\slashed{p}_4) \, .
\end{equation}

Strictly speaking, scattering amplitudes for processes such as that of
eq.~\eqref{eq:process} cannot be expressed in terms of just the set
$\{s_{12}, s_{23}, s_{34}, s_{45}, s_{51}, \text{tr}_5 \}$, as this
requires removing an arbitrary little-group-dependent factor. In fact,
amplitudes depend not only on the four-momenta, but also on the
helicities of the external states.

To better represent this dependence on the helicities, we can adopt a
different set of variables, namely the two-component spinors,
$\lambda_{i}^{\alpha}$ and $\tilde\lambda_{i}^{\dot\alpha}$, with $i
\in \{1, \dots, 5\}$. Starting from the 
$2\times 2$ spinors $p_{i}^{\dot\alpha\alpha}$, which are given in
terms of the respective four-momenta as $p_i^{\dot\alpha\alpha} =
p_{i,\mu}\sigma^{\mu\dot\alpha\alpha}$, with
$\sigma^{\mu\dot\alpha\alpha}=(\mathbb{1},\vec\sigma)$ 
and $\vec\sigma$ the Pauli matrices, the two-component
spinors $\lambda_{i}^{\alpha}$ and $\tilde\lambda_{i}^{\dot\alpha}$
are defined by noting that for massless particles
\begin{equation}
  \det(\{p_{i}^{\dot\alpha\alpha}\}) = 0 \quad \Longrightarrow \quad
  p_{i}^{\dot\alpha\alpha} =
  \tilde\lambda_{i}^{\dot\alpha}\lambda_{i}^{\alpha} \, .
\end{equation}
Lowering of spinor indices is performed as
$\lambda_{i,\alpha}=\epsilon_{\alpha\beta}\lambda_i^\beta $ and 
$\tilde\lambda_{i,\dot\alpha}=\epsilon_{\dot\alpha\dot\beta}\tilde\lambda_i^{\dot\beta}$, where we make use of 
the Levi-Civita symbol
$\epsilon^{\alpha\beta}=\epsilon^{\dot\alpha\dot\beta}=-\epsilon_{\alpha\beta}
=-\epsilon_{\dot\alpha\dot\beta}=i\sigma_2$.
Invariant contractions of spinors give so-called spinor brackets, 
which we define as
\begin{equation}
  \langle i j \rangle = \lambda^\alpha_i \lambda_{j, \alpha} \quad \text{and} \quad [ij] = \tilde\lambda_{i,\dot\alpha}\tilde\lambda_j^{\dot\alpha}\, .
\end{equation}
These are related to the Mandelstam invariants in eq.~\eqref{eq:mand} through
$s_{ij}=\langle ij\rangle [ji]$. We also use longer spinor
contractions, in particular
\begin{equation}
  \langle i | j + k | i ] = \langle i j \rangle [ji] + \langle ik
    \rangle [ki] \, .
\end{equation}
Finally, we can express $\trFive$ as a polynomial in spinor
brackets as
\begin{equation}
  \trFive = [12]\langle23\rangle[34]\langle41\rangle-\langle12\rangle[23]\langle34\rangle[41] \, .
\end{equation}

\paragraph{Helicity amplitudes} We closely follow the notation and conventions of
ref.~\cite{Abreu:2020cwb}, and denote the (renormalised) amplitudes
for this process by
\begin{align}
\label{eq:def_A}
  \mathcal{M}(1_q^{h_1},2_{\bar{q}}^{h_2},3_\gamma^{h_3},4_\gamma^{h_4},5_\gamma^{h_5})
  \coloneqq e_q^3\delta_{i_1i_2}
  \mathcal{A}(1_q^{h_1},2_{\bar{q}}^{h_2},3_\gamma^{h_3},4_\gamma^{h_4},5_\gamma^{h_5})\,,
\end{align} 
where $i_1$ and $i_2$ are the color indices of the external
quarks and $e_q$ is their electric charge. We call
$\mathcal{A}(1_q^{h_1},2_{\bar{q}}^{h_2},3_\gamma^{h_3},4_\gamma^{h_4},5_\gamma^{h_5})$
the \emph{helicity amplitudes} for the process in eq.~\eqref{eq:process}, and will often 
suppress their arguments for simplicity.

Helicity amplitudes satisfy relations under permutations of the photon momenta
or under charge and parity conjugation. For the process in \cref{eq:process}
there are two independent helicity configurations, which we choose to be
\begin{equation}\label{eq:independentHel}
  \begin{aligned}
    \mathcal{A}_{+++}(1,2,3,4,5) &\coloneqq \mathcal{A}(1_q^+,2_{\bar{q}}^-,3_\gamma^+,4_\gamma^+,5_\gamma^+)\,,\\
    \qquad
    \mathcal{A}_{-++}(1,2,3,4,5) &\coloneqq \mathcal{A}(1_q^+,2_{\bar{q}}^-,3_\gamma^-,4_\gamma^+,5_\gamma^+)\,,
  \end{aligned}
\end{equation}
where we indexed the independent amplitudes by the photon helicities. 
We work in the 't~Hooft--Veltman scheme of dimensional regularisation, setting the
space-time dimensions to $D=4-2\epsilon$, and use the definition of 
dimensionally regularised helicity amplitudes with external quarks given in 
ref.~\cite{Abreu:2018jgq}. We perform the UV renormalisation in the
$\MSbar$ scheme, where the amplitudes admit an expansion in terms of the 
renormalised QCD coupling constant $\alpha_s$ of the form
\begin{equation}\label{eq:as-series}
  \mathcal{A} = \mathcal{A}^{(0)} + \frac{\alpha_s}{2 \pi}  \mathcal{A}^{(1)}  
  + \qty(\frac{\alpha_s}{2 \pi})^2  \mathcal{A}^{(2)}  + \ldots{}\,.
\end{equation}
The coupling $\alpha_s$ is related to the bare coupling $\alpha_s^0$ through
\begin{equation}\label{eq:renormCoupling}
  \alpha_s^{0}\mu_0^{2\epsilon}S_{\epsilon} =
  \alpha_s\mu^{2\epsilon}\left( 1-\frac{\beta_0}{\epsilon}\frac{\alpha_s}{2\pi}
  + \mathcal{O} \left(\alpha_s^2\right)\right) \, , \;\quad 
  S_\epsilon=(4\pi)^{\epsilon}e^{-\epsilon\gamma_E} \, ,
\end{equation}
where $\gamma_E$ is the Euler--Mascheroni constant, $\mu_0$ and $\mu$ are the 
dimensional regularization and renormalization scales 
(which we assume to be equal), and
$\beta_0$ is the first coefficient of the QCD $\beta$-function,
\begin{equation}
  \beta_0=\frac{11}{6} C_A - \frac{2}{3} T_F \NF\,.
\end{equation} 
Here, $C_A=\NC$ is the quadratic Casimir of the adjoint
representation of the $SU(\NC)$ group, $\NF$ is the number of massless
quarks, and $T_F = 1/2$ is the normalization of the generators of the 
fundamental representation. 
Below we will also need the quadratic Casimir of the fundamental
representation, $C_F = \frac{\NC^2-1}{2\NC}$.
The coefficients of the perturbative expansion of the renormalised amplitudes
$\mathcal{A}^{(\ell)}$ are related to their bare counterparts 
$\mathcal{A}^{(\ell)}_\mathcal{B}$,
which are coefficients in a perturbative expansion in powers of $\alpha_s^{0}$, by
\begin{equation} \label{eq:renormalization}
  \mathcal{A}^{(0)}_\mathcal{B} = \mathcal{A}^{(0)} \, , \quad
  \mathcal{A}^{(1)}_\mathcal{B} = S_\epsilon \mathcal{A}^{(1)} \, , \quad
  \mathcal{A}^{(2)}_\mathcal{B} = S_\epsilon^2 \left (\mathcal{A}^{(2)} 
  + \frac{\beta_0}{\epsilon} \mathcal{A}^{(1)} \right) \, .
\end{equation}

\begin{figure}[]
\centering
  \begin{subfigure}{0.45\textwidth}
  	\centering
  	\includegraphics[trim=0 2cm 0 0,clip,scale=0.2]{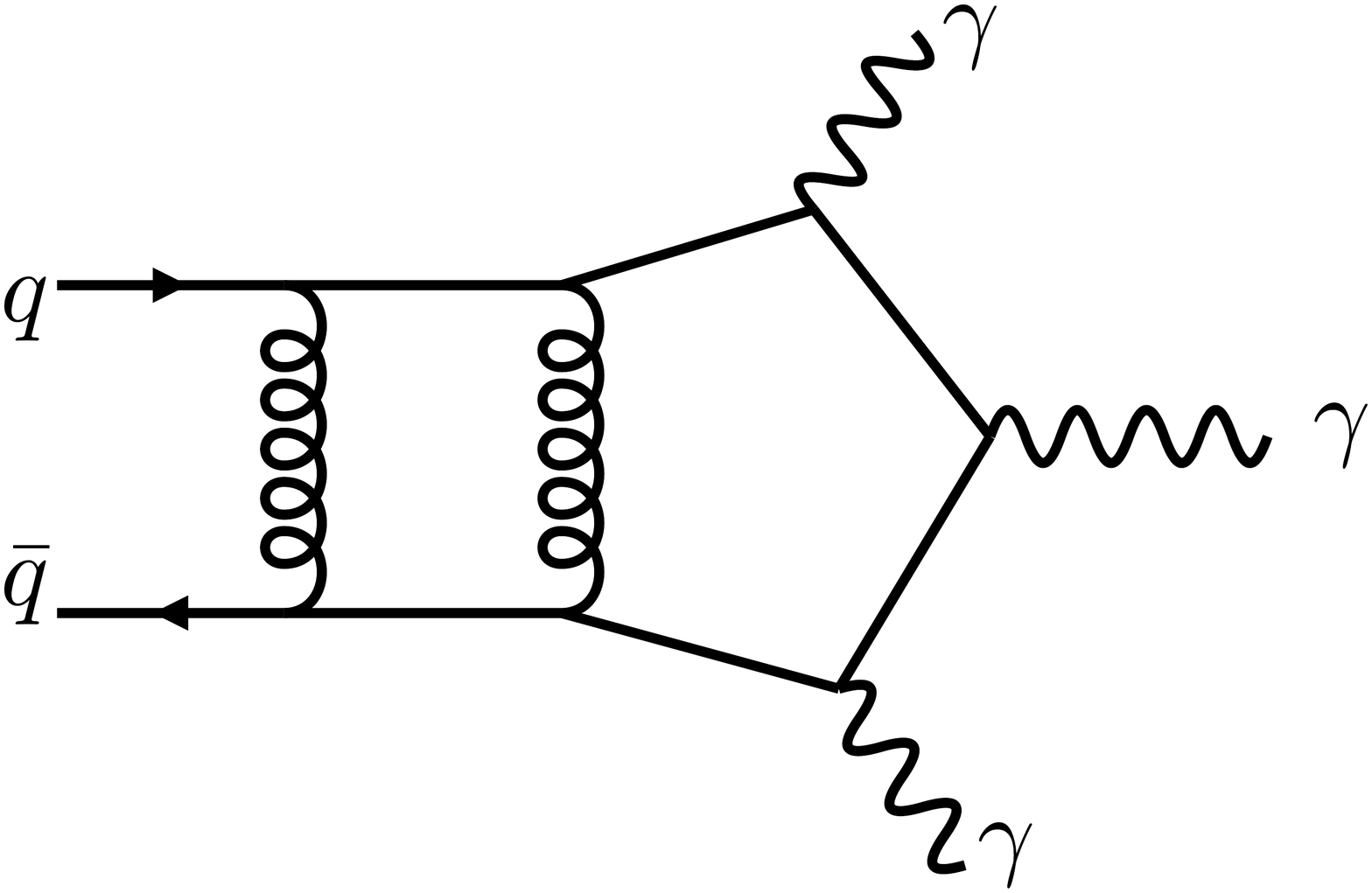}
  	\caption{$A^{(2,0)}$}
  \end{subfigure}
  \begin{subfigure}{0.45\textwidth}
  	\centering
  	\includegraphics[trim=0 1.4cm 0 0,clip,scale=0.2]{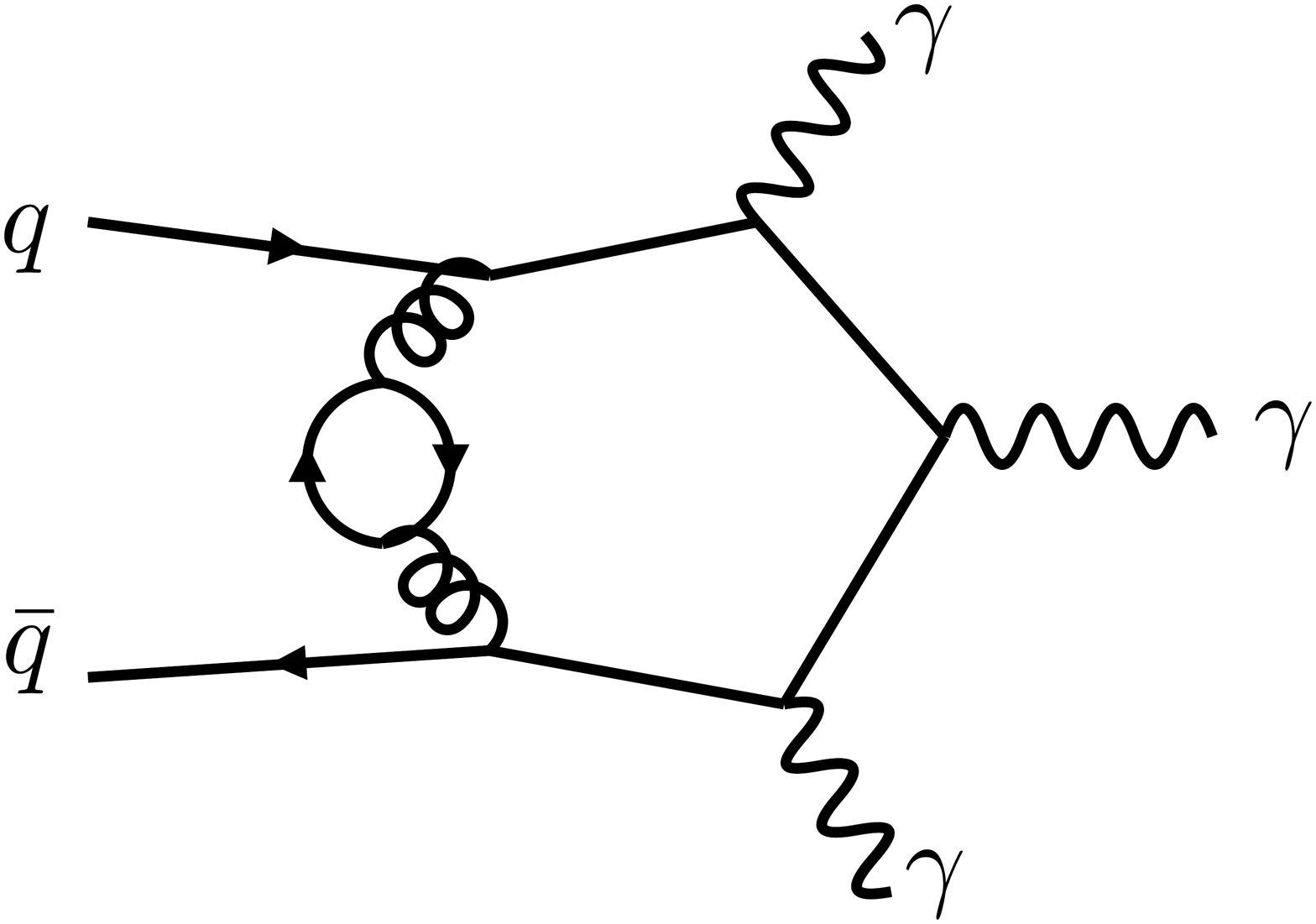}
  	 	\caption{$A^{(2,\NF)}$}
  \end{subfigure}
   \begin{subfigure}{0.45\textwidth}
  	\centering
  	\includegraphics[trim=0 5cm 0 0,clip,scale=0.2]{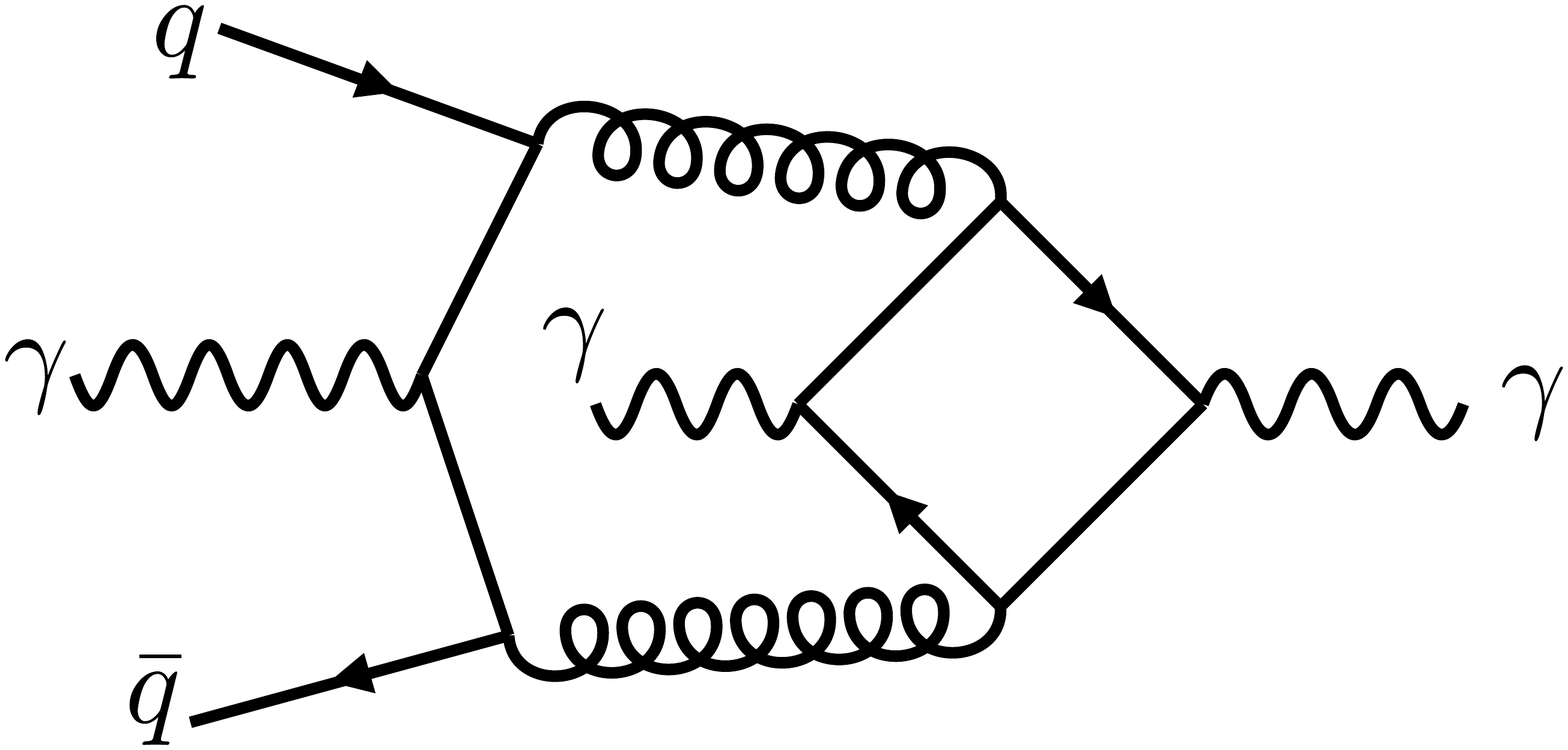}
  	\caption{$A^{(2,\NFqq)}$}
  \end{subfigure}
  \begin{subfigure}{0.45\textwidth}
  	\centering
  	\includegraphics[trim=0 5cm 0 0,clip,scale=0.2]{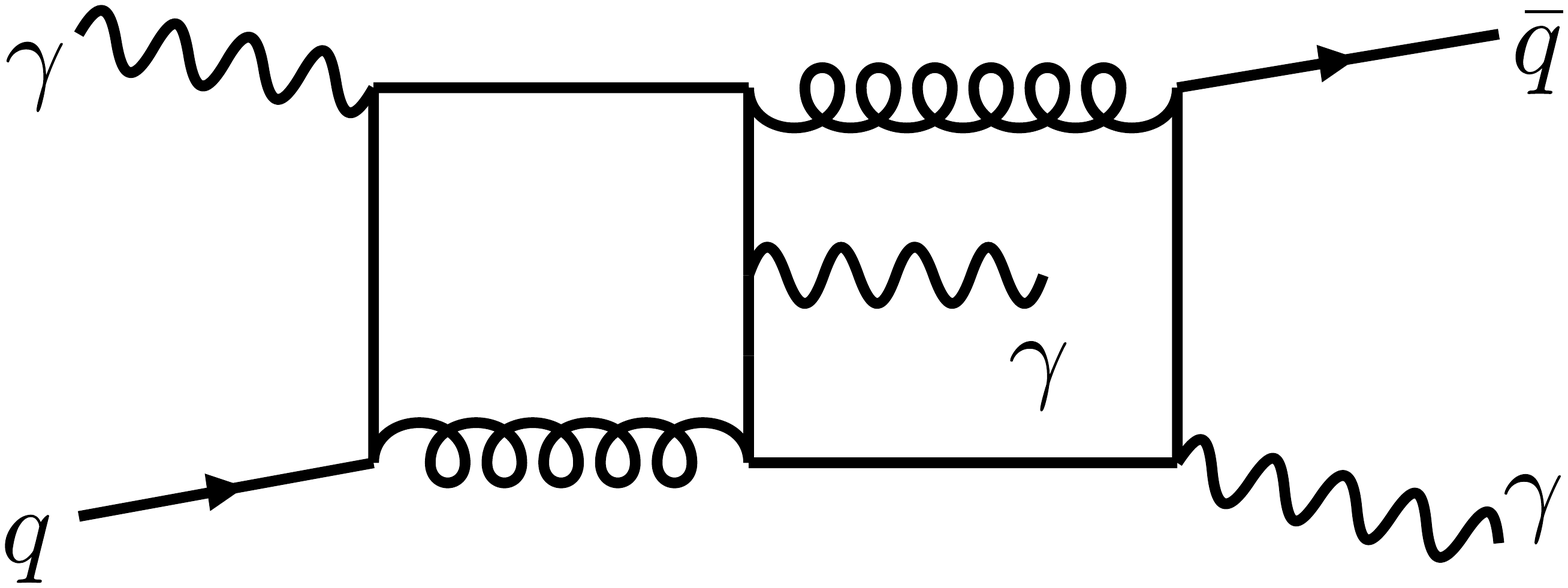}
  	\caption{ $A^{(2,1)}$}
  \end{subfigure}
\caption{Representative diagrams for the different contributions to eq.~\eqref{eq:coulourDec-gen}.}
\label{fig:representativeDiagrams}
\end{figure}

The coefficients $\mathcal{A}^{(\ell)}$ can be decomposed into individually
gauge-invariant contributions that scale differently with the number of light 
quarks $\NF$, the number of colors $\NC$
and the electric charges of the light fermions. 
Up to second order, we have \cite{Anastasiou:2002zn,Glover:2003cm}
\begin{align} \begin{split}
  \mathcal{A}^{(1)} & = C_F A^{(1)} \, ,\\
  \mathcal{A}^{(2)} & = C_F^2 B^{(2,0)} + C_F C_A B^{(2,1)} + 
  C_F T_F \NF A^{(2,\NF)} + C_F T_F \left(\sum_{f=1}^{\NF} Q_f^2\right) \, A^{(2,\NFqq)} \, ,
  \label{eq:coulourDec-gen}
\end{split}\end{align}
where $Q_f$ denotes the ratio of the charges of the light fermions running in
a closed fermion loop to the charge of the initial state quark/anti-quark pair.  
We can rearrange \cref{eq:coulourDec-gen} as
\begin{align}\begin{split}\label{eq:coulourDec}
    \mathcal{A}^{(2)} & = \frac{N_c^2}{4}\left(A^{(2,0)} -  
    \frac{1}{\NC^2}(A^{(2,0)}+A^{(2,1)})  + \frac{1}{\NC^4} A^{(2,1)} \right) \\
    & \qquad + C_F T_F \NF A^{(2,\NF)} + 
    C_F T_F \left(\sum_{f=1}^{\NF} Q_f^2\right) \, A^{(2,\NFqq)} \, ,
\end{split}\end{align}
where we have
\begin{align}
    A^{(2,0)} \coloneqq B^{(2,0)} + 2 B^{(2,1)},\qquad
    A^{(2,1)} \coloneqq B^{(2,0)}.
\end{align}
The contributions $A^{(2,0)}$ and $A^{(2,\NF)}$ involve only planar
diagrams and were previously computed in
refs.~\cite{Chawdhry:2019bji,Abreu:2020cwb,Chawdhry:2020for}.  In this
work we obtain the missing contributions $A^{(2,1)}$ and $A^{(2,\NFqq)}$. 
Representative diagrams for each contribution are
shown in figure \ref{fig:representativeDiagrams}.

The renormalised amplitudes still contain infrared (IR) poles.
They can be predicted and subtracted in a scheme-dependent way. We use the 
scheme of refs.~\cite{Catani:1998bh,Anastasiou:2002zn,Glover:2003cm},
which we denote as the Catani scheme.  
The finite remainders $\mathcal{R}$ are obtained from the 
UV-renormalised amplitudes $\mathcal{A}$ through
\begin{equation} \label{eq:remainder-def}
  \mathcal{R} \coloneqq  \mathbf{I} \mathcal{A} \, .
\end{equation}
Upon expansion of \cref{eq:remainder-def} in $\alpha_s$ and $\epsilon$,
we obtain
\begin{align} \begin{split}\label{eq:remainder2l}
    \mathcal{R}^{(0)} &= \mathcal{A}^{(0)} \, , \\
    \mathcal{R}^{(1)} &= \mathcal{A}^{(1)}+
    \mathbf{I}^{(1)}\mathcal{A}^{(0)} +\mathcal{O}(\epsilon) \, , \\ 
    \mathcal{R}^{(2)} &= \mathcal{A}^{(2)}+\mathbf{I}^{(1)}\mathcal{A}^{(1)}
    +\mathbf{I}^{(2)}{\mathcal{A}}^{(0)} +\mathcal{O}(\epsilon) \, , 
\end{split}\end{align}
where the functions $\mathbf{I}^{(1)}$, $\mathbf{I}^{(2)}$ are the coefficients
of the expansion of $\mathbf{I}$ in $\alpha_s/(2\pi)$, 
and are given in \cref{sec:catani}.\footnote{We note that our 
definition of $\mathbf{I}^{(1)}$, $\mathbf{I}^{(2)}$ 
differs by a sign from the one of ref.~\cite{Abreu:2020cwb}.} 
Up to two-loops we can expand the finite remainders into the gauge-invariant 
contributions $R^{(1)}$, $R^{(2,0)}$, $R^{(2,1)}$, 
$R^{(2,\NF)}$ and $R^{(2,\NFqq)}$ 
as in \cref{eq:coulourDec-gen,eq:coulourDec}.

For phenomenological applications, we are mostly interested in squared finite
remainders, summed over helicity and color, which we will refer to as the 
\emph{hard function}. We define it as
\begin{equation}\label{eq:hard-function-def}
  \mathcal{H} = \frac{1}{\mathcal{B}} ~ \sum_h \abs{\mathcal{R}_h}^2 \, , \qquad \mathcal{B} \coloneqq \sum_h  \abs{\mathcal{A}^{(0)}_h}^2 \, ,
\end{equation}
where the sum is over all different choices of helicities of the involved particles.
Expanding in $\alpha_s$ and in color factors, we get
\begin{align}\begin{split} \label{eq:hard-function}
    \mathcal{H}^{(1)} & = C_F H^{(1)}, \\
    \mathcal{H}^{(2)} & = \frac{N_c^2}{4}\left(H^{(2,0)} -  
    \frac{1}{\NC^2}(H^{(2,0)}+H^{(2,1)})  + 
    \frac{1}{\NC^4} H^{(2,1)} \right) \\
    & \qquad + C_F T_F \NF H^{(2,\NF)} + 
    C_F T_F \left(\sum_{f=1}^{\NF} Q_f^2\right) \, H^{(2,\NFqq)},
\end{split}\end{align}
and by definition $\mathcal{H}^{(0)} = 1$. 
We note that the hard functions are scheme dependent, with expressions in different schemes 
being related by a finite shift. We give examples of such relations 
in \cref{sec:IR-scheme-change}.


\section{Calculation}
\label{sec:calculation}

\subsection{Overview}

Our calculation is done within
the framework of two-loop numerical unitarity\
\cite{Ita:2015tya,Abreu:2017xsl,Abreu:2017hqn}. We rely on the implementation 
of the method in the program \textsc{Caravel} \cite{Abreu:2020xvt}. 

This approach starts from the observation that
the integrands of the gauge-invariant contributions $A^{(2,k)}$ 
of \cref{eq:coulourDec} all admit a decomposition of the form
\begin{equation}\label{eq:integrand}
  A(\ell)=\sum_{\Gamma}
  \sum_{i\in M_\Gamma\cup S_\Gamma}c_{\Gamma,i}
  \frac{m_{\Gamma,i}(\ell)}{\prod_{j}\varrho_{\Gamma,j}(\ell)}\,,
\end{equation} 
where the outer sum is over all distinct sets $\Gamma$ 
of inverse propagators $\varrho_{\Gamma,j}$ contributing to the amplitude
(which we call \emph{topologies}).  For each topology $\Gamma$, the
numerators $m_{\Gamma,i}(\ell)$ are polynomials in the loop momenta $\ell$ (we use
a single $\ell$ to collectively denote all loop momenta), and
they are constructed such that each $m_{\Gamma,i}(\ell)$ either corresponds
to a master integral ($i \in M_{\Gamma}$) or can be expressed as a total
derivative and therefore integrates to zero, i.e.\ it is a 
surface term ($i \in S_\Gamma$).  The coefficients $c_{\Gamma,i}$ in
\cref{eq:integrand} are unknown rational functions of the external
particles' momenta and the dimensional regulator $\epsilon$. 
In the generalized unitarity method, the
coefficients of each topology $\Gamma$ are constrained by evaluating 
(generalized) \emph{cuts}, corresponding to residues of 
\cref{eq:integrand} at values of $\ell=\hat\ell$ such that
$\varrho_{\Gamma,j}(\hat\ell)=0~\forall j$. 
The cuts of the LHS of \cref{eq:integrand} are evaluated as products of 
$D$-dimensional tree-amplitudes (which we call \emph{cut diagrams}) and 
matched to the integrand parametrisation of \cref{eq:integrand}.  
In this way, \emph{cut equations} are generated for various loop-momentum 
configurations $\hat{\ell}_k$ satisfying the cut conditions 
$\varrho_{\Gamma,j}(\hat\ell_k)=0$. Given a large enough sample of 
$\hat{\ell}_{k}$, the coefficients $c_{\Gamma,i}$ in \cref{eq:integrand}
can be determined as (numerical) solutions to the cut equations.
Our computation is based on the public implementation of the two-loop numerical
unitarity method in \textsc{Caravel} \cite{Abreu:2020xvt}, which we extended
to account for the non-planar topologies appearing in \cref{eq:integrand}.
These were absent in previous applications of the two-loop numerical unitarity
approach.  In particular, we constructed the
complete set of surface terms required for the non-planar contributions in
$A^{(2,1)}$ and $A^{(2,\NFqq)}$.
We discuss this in more detail in \cref{sec:surface_terms}.

We note that to be compatible with generalized cuts in $D$ dimensions, 
we require that the numerators $m_{\Gamma,i}(\ell)$ in \cref{eq:integrand}
are polynomials in the loop momenta (i.e.\ additional denominator powers are 
not allowed).
In addition, for each $\Gamma$ the numerators $m_{\Gamma,i}(\ell)$ must 
be linearly-independent on $D$-dimensional cuts
that set all $\varrho_{\Gamma,j}$ to zero.
While for the surface terms these condition hold by construction, 
the definitions of master integrals are frequently chosen such that at least 
one of the conditions is violated.
Indeed, the basis of master integrals of 
refs.~\cite{Chicherin:2018old,Chicherin:2020oor} violates the 
second condition.
We find that the basis of ref.~\cite{Abreu:2018aqd} is therefore more 
convenient for our approach. It can easily be written in terms
of the pentagon functions of ref.~\cite{Chicherin:2020oor} 
using modern integration-by-parts 
codes, e.g.~\cite{Klappert:2020nbg,Peraro:2019svx},
which among other benefits allows for an efficient numerical evaluation
of the master integrals. The decomposition 
of \cref{eq:integrand} then leads to a decomposition of the finite
remainders in \cref{eq:remainder2l} in terms of pentagon functions,
\begin{equation}
  \label{eq:finite-remainder-as-weighted-sum}
  \mathcal{R}^{(\ell)} = \sum_i r_i h_i \, ,
\end{equation}
where the $r_i$ are rational functions of external kinematics, 
and the $h_i$ are monomials of the pentagon functions of 
ref.~\cite{Chicherin:2020oor}.
In summary, two-loop numerical unitarity gives us a way to numerically
compute the $r_i$. We can then use these numerical evaluations to
reconstruct their analytic form.

Before discussing some details of the construction of surface terms 
in \cref{sec:surface_terms}, and of the analytic
reconstruction of the finite remainders in \cref{sec:reconstruction}, we
close this brief overview of the approach with some technical comments.
Cut diagrams are generated with \texttt{qgraf}~\cite{Nogueira:1991ex}, and
arranged into a hierarchy of cuts with a private code. To match the cuts
evaluated through color-ordered tree amplitudes to the amplitude definitions in
\cref{sec:notation} we employ the unitarity-based colour decomposition
of refs.~\cite{Ochirov:2016ewn,Ochirov:2019mtf}.  We determine the
$\epsilon$-dependence of cut diagrams originating from the state sums in the
loops through the dimensional reduction method developed in
refs.~\cite{Anger:2018ove,Abreu:2019odu,Sotnikov:2019onv}. 
This allows us to perform the entire calculation with six-dimensional states 
only.

\subsection{Surface Terms}
\label{sec:surface_terms}

The two-loop numerical unitarity framework 
\cite{Ita:2015tya,Abreu:2017xsl,Abreu:2017hqn}
builds on the parametrisation of the integrand
as in \cref{eq:integrand}.
A crucial step in this procedure is the determination
of a basis of surface terms that integrate to zero.
In this section we present the method we used to construct
surface terms for the non-planar topologies.

The construction of surface terms starts from the observation that
total derivatives of Feynman integrands integrate to zero in dimensional 
regularisation \cite{Tkachov:1981wb,Chetyrkin:1981qh,Laporta:2000dsw}. That is, we can construct surface terms 
from
\begin{align}
\label{ibp}
\int \text{d}^D\ell_1 \cdots \text{d}^D\ell_L
 \frac{\partial}{\partial \ell_a^\mu}\frac{v_a^\mu(\ell)}{\varrho_1\cdots\varrho_N}=0\,.
\end{align}
For an arbitrary vector $v_a^\mu$, the equation above will generate surface terms
involving integrals that are not relevant for the amplitudes we are computing,
either because they have numerators of too high degree or because they have propagators
raised to too high powers. In order to have an efficient construction of the
surface terms required for the decomposition in \cref{eq:integrand},
it is beneficial to construct a minimal set of surface terms, which contain
specific propagator powers and numerators whose polynomial degree is limited
by the interactions of the underlying process.

While the numerator power in the surface terms is easy to control (because
derivatives do not increase it), care must be taken with the power of the propagators.
Propagator powers in surface terms can be controlled by requiring that the
vector $v^\mu_a$ in \cref{ibp}
satisfies  \cite{Gluza:2010ws}
\begin{align} \label{eq:propdoubling}
\sum_{a,\mu} v^\mu_a(\ell) \frac{\partial \varrho_i}
{\partial \ell^\mu_a}=f_i(\ell) \varrho_i \,,
\quad \forall i\,.
\end{align}
where the unknowns $f_i$ and $v^\mu_a$ are polynomials in loop momenta.
We call such vectors unitarity-compatible integration-by-parts (IBP) 
generating vectors, or simply IBP-generating vectors.
The mathematical structure of \cref{eq:propdoubling} is well known 
and defines the vectors $v^\mu_a$ and the $f_i$ to be elements of a 
syzygy module.
 
Solving \cref{eq:propdoubling} allows us to construct a minimal set of surface terms. 
In practice, however, obtaining analytic expressions for the surface terms 
can be challenging, both due to the difficulty of 
solving \cref{eq:propdoubling}
and due to the size of the final expressions
(see refs.~\cite{Schabinger:2011dz,10.1145/1993886.1993902,Chen:2015lyz,
Bohm:2017qme,Agarwal:2020dye,Wu:2023upw} for related work).
Since our goal is to reconstruct analytic expressions for the 
two-loop remainders from their numerical evaluations on a sufficient number
of phase-space points, we do not actually require analytic solutions to
\cref{eq:propdoubling}. Indeed, it is sufficient to obtain IBP-vectors 
that are analytic in the loop-momentum variables but numerical in external
momenta. Naturally, this requires solving \cref{eq:propdoubling} at each 
phase-space point, and one must therefore have an approach that is efficient.
We now present
our solution to this problem: we first discuss its formulation in embedding
space \cite{Simmons-Duffin:2012juh, Caron-Huot:2014lda}, and then discuss how
unitarity-compatible IBP vectors and surface terms are constructed at a given
phase-space point, and how we efficiently extend this to subsequent phase-space points.

\subsubsection{Embedding-Space Formalism}

We start by reviewing the formulation of Feynman integrals in embedding space.
We will observe in the next section that this formulation simplifies the
solution of the syzygy equations in \cref{eq:propdoubling}.
Formally, our goal is to discuss how momentum space can be mapped into a subset
of a projective space, commonly called embedding space. To this end, we map
each point $z^\mu$ in momentum space
into a line $Z$ of projectively equivalent points. 
In the context of a Feynman integral with $N$ external legs, this can
be done as follows \cite{Simmons-Duffin:2012juh, Caron-Huot:2014lda}.
We first define
\begin{equation}
  q_i^\mu=\sum_{j=1}^{i-1}k_j^\mu\,,\qquad 1\leq i\leq N\,,
\end{equation}
where the $k_j^\mu$ denote the external momenta
and the empty sum gives the zero vector. We then arrange the
loop and external momenta into the $(D+2)$-dimensional embedding-space
vectors
\begin{equation}\label{eq:embedVecs}
  Y_a=c_{Y_a}\left(\ell_a^\mu, (\ell_a)^2, 1\right)\,, 
\quad X_i=c_{X_i}\left(q_i^\mu,(q_i)^2, 1 \right)\,,
\end{equation}
where $c_{Y_a}$ and $c_{X_i}$ parametrise the points in embedding space that are 
projectively equivalent.
As a projective space, embedding space is equipped with a special point, commonly referred to
as the infinity point and denoted here by $X_0$,
\begin{align}
X_0=\left(0^\mu, 1, 0\right) \,.
\end{align}
We also define an inner product as
\begin{align}
  (AB)=c_A c_B \left(-2 a^\mu\cdot b_\mu+ a^2+b^2\right)\,.
\end{align}
An $L$-loop Feynman integral is then mapped into embedding space as
\begin{equation}\label{eq:embedIntegral}
  \int \left[ \prod_{a=1}^L\text{d}^D\ell_a\right] f(\ell;q)
  =\int \left[\prod_{a=1}^L \frac{d^{D+2} Y_a}{ {\rm vol}[{\rm GL}(1)]}
  \frac{\delta[(Y_a Y_a)]}{(X_0Y_a)^{D}}\right]F(Y;X)\,,
\end{equation}
where $F(Y;X)$ is implicitly defined as the image of $f(\ell;q)$ under
the embedding procedure and ${\rm vol}[{\rm GL}(1)]$ accounts for the projective
equivalence parametrised by the $c_{Y_a}$.

In this construction, we seem to have increased the number of degrees of freedom
from $D$ to $D+2$. One apparent extra degree of freedom is
removed by noting that any $z^\mu$ in momentum space is mapped to an embedding
space point $Z$ that satisfies $(ZZ)=0$, i.e., onto the light-cone in embedding space. 
Indeed, we can easily check from \cref{eq:embedVecs} 
that $(Y_a Y_a)=0$ and $(X_i X_i)=0$. The integration in embedding space in \cref{eq:embedIntegral}
is restricted to the light-cone by the delta function $\delta[(Y_a Y_a)]$,
thereby corresponding to the integration over all loop-momentum space.
Another apparent extra degree of freedom is parametrised by the non-vanishing parameters $c_{Y_a}$ and $c_{X_i}$.
To understand how it is removed in practice, we first note that, for an arbitrary embedding-space vector $Z$,
$c_Z=(X_0 Z)$. It therefore follows that
\begin{equation}\label{eq:allVariables}
  (\ell_1-\ell_2)^2=\frac{(Y_1Y_2)}{(X_0Y_1)(X_0Y_2)}\,,\quad
  (\ell_a-q_i)^2=\frac{(Y_aX_i)}{(X_0Y_a)(X_0X_i)}\,,\quad
  (q_i-q_j)^2=\frac{(X_iX_j)}{(X_0X_i)(X_0X_j)}\,.
\end{equation}
The function $f(\ell;q)$ in \cref{eq:embedIntegral} is a rational
function of these momentum-space inner products, which implies that
$F(Y;X)$ is homogeneous of degree 0 in both $c_{Y_a}$ and $c_{X_i}$,
that is
\begin{equation}
  F(Y ;X)=
  F(Y_1,\ldots,\lambda Y_a,\ldots,Y_L ;X)=
  F(Y;X_1,\ldots,\lambda X_i,\ldots,X_N)\,,\qquad
  \lambda \neq0\,.
\end{equation}
Since the $X_i$ only appear in \cref{eq:embedIntegral} within $F(Y ;X)$,
we can set $c_{X_i}=(X_0X_i)=1$. In principle we could make the same choice for
$c_{Y_a}$, but this would obscure some of the properties of the Feynman integral,
such as the fact that in dimensional regularization the integrand has a branch point at 
$(X_0Y_a)=0$~\cite{Simmons-Duffin:2012juh,Caron-Huot:2014lda,Abreu:2017ptx}.
Equivalently, we can keep the explicit dependence on the $(X_0Y_a)$, and
the division by ${\rm vol}[{\rm GL}(1)]$ ensures that the integral is well defined
despite the invariance of $F(Y ;X)$ with respect to rescalings of each $Y_a$.

The (inverse) propagators in a Feynman integral
can all be written in terms of the expressions given in \cref{eq:allVariables},
where we are free to set $(X_0X_i)=1$ as discussed above. For
two-loop five-point non-planar integrals it is also convenient to define 
another combination of the form
\begin{align}\label{eq:propNonPlanarEmbedding}
(\ell_1-\ell_2+k_i)^2=\frac{r_{AB}Y_1^A Y_2^B}
{(X_0Y_1 )(X_0Y_2)}\,,
\end{align}
where $r_{AB}$ is a matrix that depends only on external kinematics.
That is, the components of $r_{AB}$ can be written in terms of the $(X_iX_j)$
and are independent of $Y_1$ and $Y_2$.
The denominator $(\ell_1-\ell_2)^2$ is a
particular case of \cref{eq:propNonPlanarEmbedding}.
We denote the set of propagator denominators in a two-loop 
five-point non-planar  integral as
\begin{equation}\label{eq:propagators}
{\cal P} = \big\{ \left( X_1 Y_1 \right) \,,  
\left( X_2 Y_1 \right)\,, \ldots 
\,, (Y_1 Y_2)\,, r_{AB} Y^A_1 Y^B_2 \big\}\,,
\end{equation}
where we note that the last term is not required for planar integrals.

\begin{figure}
\centering
\includegraphics[scale=0.7]{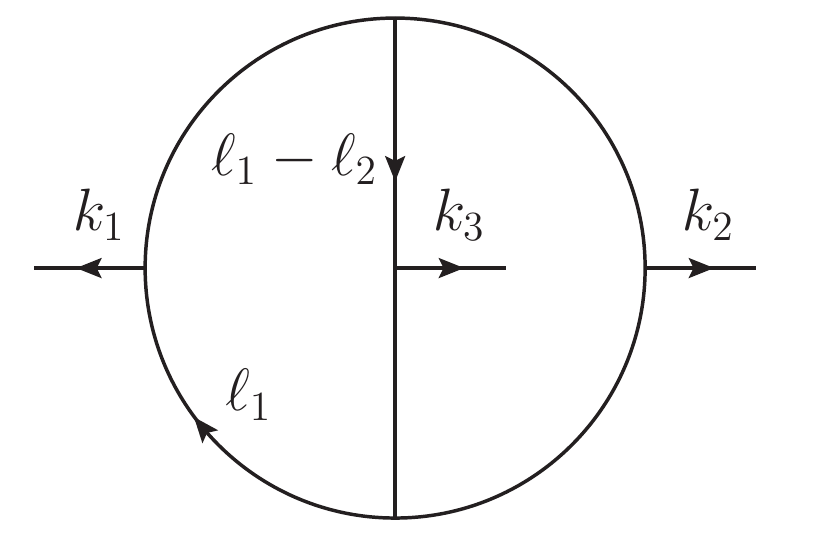} 
\caption{Non-planar diagram corresponding to the propagators in \cref{eq:propExample}.}
\label{fig:simpledoublebox}
\end{figure}

For concreteness we explicitly give an example of a simple non-planar 
diagram in embedding space. The propagators of the diagram in 
\cref{fig:simpledoublebox} take  the form
\begin{align}\begin{split}\label{eq:propExample}
\varrho_1&=\ell_1^2=\frac{(X_1Y_1)}{(X_0Y_1)}\,,\quad \varrho_2=(\ell_1-k_1)^2=\frac{(X_2Y_1)}{(X_0Y_1)}\,,\quad 
\varrho_3=(\ell_2+k_3)^2=\frac{(X_3Y_2)}{(X_0Y_2)}\,,\\
\varrho_4&=(\ell_2+k_2+k_3)^2=\frac{(X_2Y_2)}{(X_0Y_2)}\,,\quad
\varrho_5=(\ell_1-\ell_2)^2=\frac{(Y_1Y_2)}{(X_0Y_1)(X_0Y_2)}\,,\\
\varrho_6&=(\ell_1-\ell_2-k_3)^2=\frac{1}{(X_0Y_1 )(X_0Y_2)}\left[ (Y_1Y_2)+ (X_0Y_1)(X_3 Y_2) - \right.\\
&\left. \qquad\quad (X_0Y_1)(X_1 Y_2)+(X_0Y_2)(X_1 Y_1)-(X_0Y_2)(X_3Y_1)+(X_0Y_1)(X_0Y_2)(X_1X_3)\right] \,,
\end{split}\end{align} 
where $\varrho_{6}$ has the form of \cref{eq:propNonPlanarEmbedding} and we have written
the components of the respective matrix $r_{AB}$ explicitly in terms of the $(X_iX_j)$.

\subsubsection{IBP-Generating Vectors In Embedding Space}

Let us now discuss how IBP relations arise in embedding space, following previous work on the subject~\cite{Caron-Huot:2014lda,Bern:2017gdk}. 
To begin, we reformulate \cref{ibp} in embedding space as\footnote{
  In practice, it can be more convenient to replace the
$\delta\left[ \left(Y_b Y_b\right) \right]$ by $1/\left(Y_b Y_b\right)$
and modify the integration contour to encircle the pole at 
$\left(Y_b Y_b\right)$. This perspective can be taken to more easily derive some
of the properties of IBP vectors in embedding space.
}
\begin{align}\label{eq:embeddingIntegrand}
\int \prod_a \left[\frac{d^{D+2} Y_a}{ {\rm vol}[{\rm GL}(1)]}\right]  \times  \frac{\partial}{\partial Y_c^C}  
\left[\frac{ V_c^C }{\varrho_1\cdots \varrho_N}\,
\prod_b \frac{\delta\left[ \left(Y_b Y_b\right) \right]}{\left(X_0 Y_b\right)^D} \right] = 0 \,,
\end{align}
where the $\rho_i$ are now understood as their embedding-space expressions.
The IBP-generating vectors are denoted $V_c^C$, where $C$ runs over
the $D+2$ dimensions of an embedding-space vector. A vector $V_c$ that generates
linear relations between Feynman integrals in embedding space must satisfy
a number of non-trivial properties.
First, the components of $V_c^{C}$ must be homogeneous functions in $Y_c$
of degree one in order to compensate for the scaling of the partial 
derivatives. More precisely, they are required to be rational functions 
with denominators given by products of the factors $(X_0 Y_a)$, so that no 
further propagator poles are introduced.
Second, it can be verified that
\begin{align}\label{eq:vanishIntegrand}
  \frac{\partial}{\partial Y_c^C}  
\left[\frac{ F_c Y_c^C }{\varrho_1\cdots \varrho_N}\,
\prod_b \frac{\delta\left[ \left(Y_b Y_b\right) \right]}{\left(X_0 Y_b\right)^D} \right] = 0 \,,
\end{align}
for any homogeneous $F_c$ that is degree 0 in $Y_c$.
As such, any IBP-generating vector $V_c$ generates the same surface term as another
vector obtained from it by translations in the $Y_c$ direction.
Finally, any valid IBP vector must only generate relations between
Feynman integrals. We therefore impose that any terms containing derivatives of
the delta function that arise from expanding the derivative in
\cref{eq:embeddingIntegrand} should cancel. This can be achieved by requiring that
\begin{align}\label{eq:constraintQuadric}
 V_c^C \frac{\partial}{\partial Y_c^C} (Y_b Y_b)=
 F^{\rm quadric}_b \, (Y_b Y_b)\,,    \qquad  \forall b \,,
 \end{align}
i.e.\ by requiring that the $V_c$ generate translations along the light cone.
It turns out that many of the solutions to \cref{eq:constraintQuadric} can be
discarded as they lead to vanishing surface terms.
Indeed, it is easy to see that a subset of the solutions is given by $V_c$ such
that $V_c=\frac{1}{2} F_c^{\rm quadric} Y_c$. By comparison with
\cref{eq:vanishIntegrand}, we see that such solutions give a vanishing surface
term.
We can use this observation to simplify IBP-vector construction. Specifically,
if we have an IBP vector $V_c$ for which the associated $F_c^{\rm quadric} \ne
0$, then the same surface term is generated by a vector
$V'_c=V_c-\frac{1}{2} F_c^{\rm quadric} Y_c$. It follows that
the new vector $V'_c$ satisfies equation
\eqref{eq:constraintQuadric} with $F^{\rm quadric}_b=0$. Therefore, the
solutions of \cref{eq:constraintQuadric} with $F^{\rm quadric}_b=0$ generate
the full set of surface terms and it is sufficient to require that
the IBP vectors $V_c$ satisfy
\begin{align}\label{eq:constraintQuadricSimp} 
  V_c^C \frac{\partial}{\partial Y_c^C} (Y_b Y_b) = 0\,,    \qquad  \forall b \,.  
\end{align}

Within the two-loop generalised unitarity framework, we
also require that IBP-generating vectors
are unitarity compatible, that is that they satisfy \cref{eq:propdoubling}.
In embedding space, the analogous condition is that the exponents of the 
denominator factors $(X_i Y_a)$ and $\big(r_{AB} Y^A_1 Y^B_2\big)$ are not 
increased \cite{Bern:2017gdk}.
For propagators that depend on a single $Y_a$, 
the conditions read,
\begin{align}\label{eq:constraintP}
 V_c^C\frac{\partial}{\partial Y_c^C} \left(X_i Y_b\right)  
  = F_i \,\left( X_i Y_b\right)\,,
\end{align}
and for the propagators that depend on both loop momenta
\begin{align}\label{eq:constraintNP}
 V_c^C \frac{\partial}{\partial Y_c^C} \left( r_{AB} Y_1^A Y_2^B \right)
  = F_r \left( r_{AB} Y_1^A Y_2^B \right)  \,,
\end{align}
which also covers the case of the propagator $(Y_1Y_2)$.

In summary, we must construct IBP-generating vectors satisfying 
\cref{eq:constraintQuadricSimp,eq:constraintP,eq:constraintNP}. These
constrain the vectors $V_a^A$ and the polynomials $F_i$ and $F_r$ to be elements of 
the syzygies of a module defined by 
\begin{align}\begin{split}\label{eq:syzEmbedding}
 \left( V_a  Y_a\right) &=0\,, \quad  \forall a\,,\\
 \left( V_a  X_i\right) &= F_i \left(X_i Y_a\right)\,, \quad  
 \text{for }\left(X_i Y_a\right) \in {\cal P} \,,\\
 r_{AB} \left[ V_1^A Y^B_2  + V_2^B Y^A_1 \right]   &
 = F_r \left( r_{AB} Y_1^A Y_2^B\right)\,, 
	\quad  \text{for } (Y_1Y_2) \text{ and }\left(r_{AB} Y_1^A Y_2^B\right) \in {\cal P} \,,
\end{split}\end{align}
where ${\cal P}$ was defined in \cref{eq:propagators}.
The unknowns in these equations are the $V_a^A$, $F_i$ and $F_r$. 
From now on, for simplicity we set $(X_0Y_a)=1$ and require the solutions to be polynomials in
the $(X_iY_a)$  and $(Y_1Y_2)$, keeping in mind that those factors can be
reinstated by requiring that quantities have the correct homogeneous degree. 

\subsubsection{Computing IBP-Generating Vectors} 
\label{sec:solvingSyzygy}

We now turn towards solving the syzygy equations \eqref{eq:syzEmbedding}.
In short, our strategy is to construct a degree-bounded generating set of
solutions on a single phase-space point and then use the shape of this
set to compute a generating set of vectors on further phase space
points, via linear algebra.
We start by expressing the IBP vectors in terms of the momenta in the problem,
\begin{align}\label{eq:param} 
V_a= \sum_{i=0}^N v^i_a X_i + v^Y_a Y_{b\neq a},
\end{align}
where we took into consideration \cref{eq:vanishIntegrand} and the
surrounding discussion to remove the $Y_a$ direction.
We can then insert this parametrisation into \cref{eq:syzEmbedding}
and re-express these equations as
\begin{align}\label{eq:syzM}
M_{IJ} \,w^J = 0 \,,\quad  w^J=\{v^i_1,v^Y_1,v^i_2,v^Y_2,F_i,F_r\} \,,
\end{align}
where the vector $w^J$ collects all unknowns in the problem (while
it contains a single $v_1^Y$ and $v_2^Y$, there are several $v^i_1$ and $v^i_2$,
several $F_i$ and two $F_r$).
The direct solution of \cref{eq:syzM} is in general very challenging.
Since we are only interested in a subset of solutions that generates
sufficiently many surface terms to construct the decomposition in
\cref{eq:integrand}, which is degree bounded by the theory, we instead solve the
problem for a fixed polynomial degree \cite{Schabinger:2011dz,Abreu:2017hqn,Wu:2023upw}. To
make this degree manifest, we consider the ansatz
\begin{align}\label{eq:linAlg}
w^J=\sum_{|\vec m| \le Q } w^J_{\vec m} y^{\vec m} \,,\qquad 
y=\{ \left(X_1 Y_1 \right),  \ldots, \left(X_1 Y_2 \right), \ldots , \left(Y_1 Y_2 \right) \} \,,
\end{align}
where the $w^J_{\vec m}$ are $Y$-independent rational functions of the external kinematics,
i.e.\ of the $(X_iX_j)$, that multiply associated monomials of the $y_k$
variables, 
\begin{align}\label{eq:monomials}
y^{\vec m} := \prod_k y_k^{m_k} \,,
\end{align}
where the multi-index $\vec m$ specifies the exponents of the monomial.
The total degree of a monomial $y^{\vec m}$ is denoted by $|\vec m|$,
\begin{align}\label{eq:absdegree}
 \quad |\vec m|:=\sum_k m_k \,.
\end{align} 
Inserting \cref{eq:linAlg} into \cref{eq:syzM}  
one obtains a linear system for the coefficients $w^J_{\vec m}$
allowing us to solve the syzygy problem with a degree bound $Q$ as a 
linear algebra problem.

We can however further simplify the problem. Indeed, we note that
the matrix $M_{IJ}$ is linear in the $y_k$, that is
\begin{align}\label{eq:decomposition}
M_{IJ}=M_{IJ}^{[0]}+M_{IJ}^{[1]}\,,\qquad
M_{IJ}^{[1]}=\sum_{k}M_{IJk}^{[1]} y_k\,.
\end{align}
It is interesting to note that, as the $X_i$ are linearly independent, $M_{IJ}^{[1]}$ is independent of external kinematics, i.e.,
the $M_{IJk}^{[1]}$ are matrices of rational numbers. 
We can also decompose the ansatz of \cref{eq:linAlg} into leading
and subleading $y$ contributions,
\begin{align}\label{eq:decompositionSol}
w^J &=w^J_{\rm max} + w^J_{\rm rem}\,,\qquad
w^J_{\rm max} = \sum_{|\vec m|=Q} w_J^{\vec m}  y^{\vec m} \,.
\end{align}
It then follows that
\begin{align}\label{eq:leadingSyz} 
M_{IJ}^{[1]} \,w^J_{\rm max} &=0 \,,
\end{align}
which is a simpler syzygy problem than \cref{eq:syzM} as it is
homogeneous and independent of kinematics. For $w^J$ to be a solution to \cref{eq:syzM},
it is necessary that its maximal degree piece solves \cref{eq:leadingSyz}.

To solve \cref{eq:leadingSyz} we use the methods implemented in
\textsc{Singular}~\cite{DGPS}, which yield a generating set of solutions,
with each generator homogeneous of a given degree. That is, we have 
the generating set,
\begin{align}\label{eq:homSyz}
\left\{ w_{k}^{J,[q_k]}(\vec y) \right\}_{k=1,\,...\,k_{\rm max}} \,,
\end{align}
where $q_k$ is the degree of the homogeneous generator $w_{k}^{J,[q_k]}$.
A basis of solutions with degree $q$ to \cref{eq:leadingSyz} is obtained from
linear combinations of the generating set of \cref{eq:homSyz} with coefficients
that are homogeneous polynomials of degree $q-q_k$.

We now use the homogeneous syzygies of \cref{eq:leadingSyz} 
to construct solutions to the full syzygy
problem of \cref{eq:syzM}. Since we require solutions only up to a maximal
polynomial degree we again consider solutions up to a given polynomial degree
$Q$.
To this end, 
we consider a parametrization of a syzygy with degree $q$ that consists of
\begin{enumerate}
\item a maximal degree piece of degree $q$ constructed from the generators
\cref{eq:homSyz}. This consists of combinations of the generators multiplied
with polynomials of suitable polynomial degree, and,
\item a generic polynomial-valued vector with degree less than $q$.
\end{enumerate}
More precisely, such a syzygy can be written as
\begin{align} 
v^{I,[q]}(\vec y) = 
\sum_{k,\,\, |\vec m_k| + q_k=q } b_{k,\vec m_k}
\,   y^{\vec m_k}\, w_{k}^{I,[q_k]}(\vec y)
+ \sum_{|\vec m|<q} b^I_{\vec m} y^{\vec m}\,, 
\end{align} 
for $y$-independent coefficients $b_{k,\vec m_k}$ and $b^I_{\vec m}$. The
monomials $y^{\vec m}$ and $y^{\vec m_k}$ are bounded by the degree 
constraints $|\vec m|<q$ and $|\vec m_k|<q$.
Inserting this ansatz into \cref{eq:syzM} then yields linear equations for
$b_{k,\vec m_k}$ and $b^I_{\vec m}$. 
In practice, we observe that this approach of combining maximal degree 
solutions
with a parametrization of lower degree terms significantly reduces the
size of the linear system compared to starting from
the generic parametrisation of \cref{eq:linAlg}.
In this way, working degree by degree we can construct a basis of the 
solutions up to degree $Q$.
Once we have constructed this basis of the solutions it is easy to find a 
subset of the basis elements which generate the module up to degree $Q$. 
Working degree by degree, one simply removes basis elements that are 
polynomial combinations of lower degree generators using linear algebra.

At this point we have obtained a generating set of the solutions of
\cref{eq:syzM} up to polynomial degree $Q$ at a given numerical phase-space
point.
We now use this information to streamline the construction of IBP
vectors on further phase-space points.
We start with the generating set we have obtained
and express their components as a
linear combination of monomials
\begin{align}
v^I(\vec y)=\sum_{\vec{m}} r^I_{\vec{m}} y^{\vec{m}},
\end{align}
where we sum over all monomials such that the associated $r^I_{\vec{m}}$ are
non-zero.
We then replace the numerical coefficients $r^I_{\vec{m}}$ by parameters
$b^I_{\vec{m}}(\vec p)$ that explicitly depend on the external kinematic 
point $\vec p$, in order to obtain \textit{skeleton} vectors
\begin{align}
  \label{eq:skeletonDefinition}
v^I_\text{skel} (\vec y)=\sum_{\vec m} b^I_{\vec{m}}(\vec p) {\vec y}^{\vec{m}}.
\end{align}
We finally insert these skeleton vectors into \eqref{eq:syzM}, 
and obtain a linear system of equations that constrain
the $b^I_{\vec{m}}(\vec p)$. Importantly, these linear systems are analytic in
the external kinematics and so can be solved phase-space point by phase-space
point.
These equations may be linearly dependent and not uniquely determine the
$b^I_{\vec{m}}(\vec p)$. We discard linearly dependent equations, and choose a
set of $b^I_{\vec{m}}(\vec p)$ to set to zero so that the solution is unique.
By making use of these skeleton vectors instead of fully general vectors, we
obtain equations for a minimal set of monomial coefficients. In practice, the
largest linear systems that we encounter when solving for the $b^I_{\vec{m}}$ of
\cref{eq:skeletonDefinition} are approximately
$2000\times 2000$.
Finally, we note that for simpler case we find it efficient to solve these
linear systems analytically.

\subsubsection{Computing Surface Terms}

Now that we have constructed a set of IBP generating vectors up to degree $Q$,
we use them to construct a collection of surface terms.
Recalling \cref{eq:integrand}, our goal is to construct a set of
linearly-independent surface terms $S_\Gamma$ for a given topology $\Gamma$ that
is sufficiently large for the power counting of our theory.
This set of surface terms is a subset of the $m_{\Gamma,i}$ of
\cref{eq:integrand} and they are polynomials in the loop momentum components.
The maximum degree of these polynomials is known a priori, since it is determined
by the theory describing the scattering process, and as such it is easy to
construct a basis of $\text{span}(M_\Gamma\cup S_\Gamma)$. As a basis of master
integrals is also known
\cite{Chicherin:2018old,Abreu:2018aqd,Chicherin:2020oor}, we have sufficient
information to construct a linearly-independent set of surface terms.

To construct a single surface term, we start with an IBP-generating vector and a monomial
of the $y_k$ variables. We take their product, and use this as the vector
$V_c^C$ in \cref{ibp}. Expanding out the derivative leads to a surface term in
embedding space that is easily mapped to one in momentum space by identifying
all $y_k$ variables with their momentum space counterparts.
In this way, taking all pairs of IBP-vectors and monomials of $y_k$ variables
that satisfy our power counting bounds, we construct a set of surface terms.
Many of the surface terms obtained in this way are linearly dependent and can
be discarded.
As our construction of the IBP-generating vectors was only performed up to
degree $Q$, the set of surface terms generated in this way may not be a basis
of $\text{span}(S_\Gamma)$. 
In practice, to solve this problem, we repeat the procedure described in 
\cref{sec:solvingSyzygy}, increasing the value of $Q$ in
order to construct more IBP generating vectors until our set of surface
terms is a basis of $\text{span}(S_\Gamma)$.

\subsection{Analytic Reconstruction In Spinor Helicity Formalism}
\label{sec:reconstruction}

As noted above, the two-loop numerical unitarity approach allows us to 
numerically evaluate the coefficients $r_i$ in the finite remainder of
\cref{eq:finite-remainder-as-weighted-sum}. We now discuss
how to obtain analytic expressions using these numerical evaluations.
As suggested in e.g.\ refs.~\cite{DeLaurentis:2020qle,DeLaurentis:2022otd},
we will consider the $r_i$ to be rational functions of the
spinor-helicity variables $\lambda$ and $\tilde\lambda$ already introduced
in \cref{sec:notation}.\footnote{More
precisely, the coefficients belong to the field of fractions of the ring of
independent Lorentz invariants; see ref.~\cite[Section
2.2]{DeLaurentis:2022otd} for more details.}
The coefficients admit a least common denominator representation, which reads
\begin{equation}\label{eq:CommonDenominatorForm}
  r_i(\lambda, \tilde\lambda) = \frac{\mathcal{N}_i(\lambda, \tilde\lambda)}{\prod_j \mathcal{D}_j^{q_{ij}}(\lambda, \tilde\lambda)} \, .
\end{equation}
The exponents $q_{ij}$ are allowed to take negative values, thus denoting
numerator factors. The $\mathcal{N}_i$ and $\mathcal{D}_j$ are polynomials in
spinor brackets.
Like the amplitude, the $r_i$ are dimensionful and transform non-trivially under the 
little group. For a spinor function $\mathcal{E}(\lambda, \tilde{\lambda})$ 
we define the mass dimension $[\mathcal{E}]$ and $k^\mathrm{th}$ 
little-group weight $\{\mathcal{E}\}_k$ through
\begin{align}
  \label{eq:MassDimensionDefinition}
  \mathcal{E}(z\lambda_1, \ldots, z\lambda_n, z\tilde{\lambda}_1, \ldots, z\tilde{\lambda}_n) 
	&= z^{2[\mathcal{E}]}\,\mathcal{E}(\lambda_1, \ldots, \lambda_n, \tilde{\lambda}_1, \ldots, \tilde{\lambda}_n)\,, \\
  \mathcal{E}(\ldots, z\lambda_k, \ldots, \tilde{\lambda}_k/z, \ldots) 
	&= z^{\{\mathcal{E}\}_k}\,\mathcal{E}(\ldots, \lambda_k, \ldots, \tilde{\lambda}_k, \ldots).
    \label{eq:LittleGroupDefinition}
\end{align}
The mass dimension of an $n$-point amplitude $\mathcal{A}_n$ is 
$[\mathcal{A}_n]= 4 - n$. 
The $k^\mathrm{th}$ little-group weight of an amplitude depends on 
the $k^\mathrm{th}$-particles' helicity $h_k$,
\begin{align}
 \{\mathcal{A}\}_k= -2 h_k \, .
\end{align}
Crucially, there is by now a large collection of evidence that the
$\mathcal{D}_j$ in \cref{eq:CommonDenominatorForm} are known a 
priori. Indeed, they can be constructed from the symbol
alphabet~\cite{Abreu:2018zmy}.
In our normalisation of the amplitudes (see \cref{eq:def_A}),
they also include factors with little-group weight, such as the 
tree level amplitude if it is not zero, or the leading order of
the one-loop amplitude.
Given that the pentagon functions $h_i$ in
\cref{eq:finite-remainder-as-weighted-sum} are dimensionless and 
little-group invariant, the mass dimensions and little-group weights of 
the $r_i$ are the same as those of the amplitudes. Since the
${\cal D}_j$ are known,  the mass dimensions and little-group weights of
the numerators $\mathcal{N}_i$ are easily determined.

In a nutshell, the reconstruction procedure amounts to constructing
and fitting an ansatz for the $\mathcal{N}_i$ in terms of spinor variables.
An important feature of spinor-helicity-based reconstruction methods is 
that the ansatz directly captures the little-group and mass-dimension 
properties of the coefficients. 
This is in contrast to Mandelstam-based reconstruction methods,
where the $r_i$ are normalized with a (dimensionful) phase factor to be
little-group invariant, and split into parity even and odd parts.
This difference has an important practical consequence, as the common 
denominator form of the $r_i$ often simplifies in spinor variables. 
For example, the symbol alphabet may factorize in spinor variables. 
This can lead to denominator factors that cancel further against 
the coefficient's numerators, lowering the mass dimensions of the latter,
which thus require simpler ans\"atze in the reconstruction procedure.

To begin constructing our ansatz, we specify the set of denominator factors 
${\cal D}_j$ in eq.~\eqref{eq:CommonDenominatorForm}. 
As already alluded to, these denominator
factors can be constructed from the symbol alphabet. 
In Mandelstam variables they correspond to the subset of so-called
even letters, which describe the collection of singular surfaces 
associated to the master integrals. 
When written in terms of spinor variables, these surfaces may branch. 
That is, alphabet letters that appear irreducible when written in terms of 
Mandelstam variables may further factorize in spinor space. 
Beyond alphabet letters, the ${\cal D}_j$ also include rational functions
appearing in tree-level or one-loop amplitudes.
We therefore take these functions and the
irreducible factors of the symbol alphabet in spinor space as the set of
expected denominator factors.
This analysis was already performed in ref.~\cite{DeLaurentis:2022otd} 
for the amplitudes considered in this paper, and it was found that the
expected set of denominator factors contains $35$ elements. 
It can be expressed as
\begin{equation}\label{eq:codimension-one-poles}
  {\cal D} = 
  \big\{\langle ij \rangle, \, [ij] \,\,:\,\, 1 \le i < j \le 5\big\} \cup \left(\bigcup_{\sigma \in Z_5} \sigma \circ \big\{\langle 1|2+3|1], \langle 1|2+4|1],  \langle 1|2+5|1] \big\}\right),
\end{equation}
where the elements of $Z_5$ are the cyclic permutations acting on the 
momentum indices associated to each spinor.
Note that, despite its presence in the alphabet we do not include
$\mathrm{tr}_5$ in $\mathcal{D}$, as it is expected to cancel in the finite
remainder (see e.g.~refs.~{\cite{Abreu:2018aqd,
Chicherin:2018yne,Chicherin:2020umh}).

In the following, it will be useful to consider the zero set associated to 
various denominators. We denote the algebraic variety corresponding to the
common zero set associated to a list of denominator factors 
$\{ {\cal D}_{i_1} \,,\ldots\,, {\cal D}_{i_n} \}$ by
\begin{equation}
  \label{eq:ideal} 
V( {\cal D}_{i_1} \,,\ldots\,, {\cal D}_{i_n} ) \, .
\end{equation}

\paragraph{Denominator exponents}

Our first task is to determine the denominator exponents $q_{ij}$ in
eq.~\eqref{eq:CommonDenominatorForm}. We will use the standard
technology of univariate-slice reconstruction~\cite{Abreu:2018zmy}, combined 
with the approach introduced in ref.~\cite{PageSAGEXLectures}, based on
all-line BCFW-shifts~\cite{Britto:2005fq,Risager:2005vk,Elvang:2008vz}. Its
application to functions of Mandelstam variables was discussed
in ref.~\cite{Abreu:2021asb}. Here we review the procedure and discuss its
generalization to functions of spinor variables.

In order to determine the $q_{ij}$, we must choose a set of
curves in phase space that intersect each surface $V(\mathcal{D}_j)$ at a
generic point at least once and are not contained in any of the
$V(\mathcal{D}_j)$.
We will achieve this with two BCFW shifts, one holomorphic and one
anti-holomorphic, following the all-line shift approach of 
ref.~\cite{Abreu:2021asb} which we now review.
Let us begin with a generic, numeric momentum-conserving configuration of
spinors $\{\lambda_1,\tilde{\lambda}_1,...,\lambda_5,\tilde{\lambda}_5\}$.
We make an all-line holomorphic shift by adjusting every $\lambda$ spinor in a
way which is proportional to a common reference spinor $\eta$. That is we shift
\begin{align}\label{eq:PureHolomorphicShift}
\lambda_i \rightarrow \lambda_i + t c_i \eta \, , \quad \tilde\lambda_i \rightarrow \tilde\lambda_i \, .
\end{align}
Here, we introduce unknowns $c_i$ that we use to ensure that the shifted
kinematics satisfy momentum conservation. Specifically, we choose the $c_i$ to
satisfy
\begin{align}
\sum_{i=1}^5 c_i \tilde{\lambda}_i = 0\,.
  \label{eq:AllLineShiftConstraint}
\end{align}
This linear equation does not have a unique solution. Nevertheless, any 
solution with $c_i \neq 0$ will guarantee a non-trivial shift that
satisfies momentum conservation. In order to ensure that a slice is generic,
we pick both the initial set of spinor variables
$\{\lambda_1,\tilde{\lambda}_1,...,\lambda_5,\tilde{\lambda}_5\}$ and an
independent subset of the $c_a$ randomly over a finite field.
The corresponding anti-holomorphic shift can be constructed by taking the 
parity conjugate.

Let us now consider how a holomorphic shift affects the set of denominator
factors $\mathcal{D}$ in \cref{eq:codimension-one-poles}. 
Firstly, the holomorphic spinor products are linear functions of $t$
\begin{align}\label{eq:AngleBracketUnderShift}
  \langle ab \rangle \rightarrow \langle ab \rangle + t \, \big(c_a \langle \eta b \rangle + c_b \langle a \eta \rangle \big) \, ,
\end{align}
whereas the anti-holomorphic spinor products remain
unchanged.
It is easy to see that the curve parametrised by $t$ intersects each of the ten
distinct codimension-one varieties $V\left(\langle ab \rangle
\right)$ as well as each of the fifteen varieties $V\left(
\langle a|b+c|a] \right)$. Nevertheless, it does not intersect any of the ten
varieties $V\left([ ab ]\right)$.
By comparing the form of the $\mathcal{D}_j$ on this univariate slice with the
denominator of $r_i$ on the slice, one can compute the exponents $q_{ij}$
corresponding to the $\mathcal{D}_j$ which are not purely anti-holomorphic (i.e.~the $[ab]$).
In order to determine the exponents of the $[ab]$ denominator factors, we repeat
the procedure on the anti-holomorphic shift.
We note that the anti-holomorphic shift must yield the same $q_{ij}$ for
$\langle a|b+c|a]$ as the holomorphic one, providing a consistency check.
Having determined the $q_{ij}$ exponents in \cref{eq:CommonDenominatorForm},
we can now determine the mass dimension and spinor weight of the numerators
$\mathcal{N}_i$, allowing us to write an ansatz for them in terms of 
spinor-helicity variables.

\paragraph{Improved Ansatz}
It is well known that the $r_i$ are not all independent rational functions,
and it is sufficient to reconstruct a basis of this space of functions.
We begin by sorting the $r_i$ by the mass dimension of their numerator
$\mathcal{N}_i$, and then apply standard linear-algebra techniques
on finite-field-valued evaluations of the $r_i$ to determine
a basis of the function space of the $r_i$.
We therefore express the pentagon-function coefficients as
\begin{equation}
  r_i = \tilde{r}_j M_{ji}  \, ,
  \label{eq:pentagon_function_coefficients_projected_to_basis}
\end{equation}
where $M_{ji}$ is a rectangular matrix with $M_{ji} \in \mathbb{Q}$.
This observation allows us to reduce the number of functions to reconstruct.

A further important simplification comes from the observation that
the rational coefficients in a scattering amplitude
are less naturally expressed in common-denominator form, and should in fact be
cast in some partial-fractions representation of the form
\begin{equation}\label{eq:PartialFractionedForm}
  \tilde{r}_i = \sum_k \tilde{r}_{ik}, \qquad \text{with} \qquad \tilde{r}_{ik} = \sum_k \frac{\mathcal{N}_{ik}}{\prod_j \mathcal{D}_j^{q_{ijk}}} \, ,
\end{equation}
where the $\mathcal{N}_{ik}$ are polynomials in spinor brackets and 
the $q_{ijk}$ are integer exponents that determine the exact 
partial-fractions decomposition.
For analytic reconstruction approaches, this means that in general 
a common-denominator ansatz is unnecessarily large.
However, systematically constructing a compact partial-fractions 
ansatz for a generic amplitude remains a challenging task 
(see refs.~\cite{DeLaurentis:2019bjh,
DeLaurentis:2022otd, DeLaurentis:2022knk} for recent progress in this area).
A practical solution to this issue is to explore several possible 
partial-fractions decompositions, determine the size of the ansatz of the 
$\mathcal{N}_{ik}$ for each of them, and declare that a suitable (if not
optimal) decomposition has been found when the ansatz is small enough for the 
$\mathcal{N}_{ik}$ to be reconstructed. For the amplitudes we are
concerned with in this paper, however, we were able to identify
a very convenient partial-fractions decomposition that we now discuss.

To construct compact ans\"atze for the rational functions, we build
upon the observation that for several one-loop five-point massless amplitudes
all poles of the
form $\langle c | a+b |c]^\alpha$ can be separated into different fractions
(see e.g.\ \cite{Bern:1993mq}).
This feature is related to the fact that the poles 
$\langle c | a+b |c]$ are spurious, and can be used to greatly simplify
analytic expressions for amplitudes \cite{Bern:1997sc}.
Given this observation, we work under the assumption that a one-loop-like
partial-fractions representation exists at two loops. 
The validity of this assumption will be tested during the calculation.
To take an explicit example, consider a function
whose common-denominator form reads
  \begin{equation}
    \frac{\mathcal{N}}{⟨1|2+4|1]⟨1|2+5|1]⟨2|1+5|2]⟨4|1+2|4]^2⟨4|1+5|4]^2⟨5|1+2|5]^2⟨5|1+4|5]^2}\,, \nonumber
\end{equation}
where $\mathcal{N}$ is an unknown polynomial in spinor brackets to be
determined.
The partial-fractions decomposition would then take the form
\begin{align}\begin{split}
    &\frac{\mathcal{N}_{1}}{⟨1|2+4|1]} + 
    \frac{\mathcal{N}_{2}}{⟨1|2+5|1]} + 
    \frac{\mathcal{N}_{3}}{⟨2|1+5|2]}+ 
    \frac{\mathcal{N}_{4}}{⟨4|1+2|4]^2} \\
    & \qquad + \frac{\mathcal{N}_{5}}{⟨4|1+5|4]^2} + 
    \frac{\mathcal{N}_{6}}{⟨5|1+2|5]^2} + 
    \frac{\mathcal{N}_{7}}{⟨5|1+4|5]^2} \, ,
  \label{eq:partial_fractioned_r174}
\end{split}\end{align}
where the $\mathcal{N}_{i}$ are unknown polynomials in spinor brackets.
The ans\"atze for each of them is substantially simpler than that for
the numerator  $\mathcal{N}$ in the common-denominator form, making the
reconstruction substantially more efficient.

Another layer of simplification of the reconstruction procedure comes from 
the observation that, when considering a partial-fractions representation,
many of the basis functions are {\em partially} contained within the vector
space spanned by the others.
More precisely, given a basis element $\tilde{r}_n$ expressed in the form of
eq.~\eqref{eq:PartialFractionedForm}, many of the $\tilde{r}_{n k}$ belong to
$\Span(\tilde{r}_{i\neq n})$. If we consider the reconstruction of a 
basis function $\tilde{r}_n$, this motivates us to include other basis 
functions in its ansatz, while dropping many of the terms in
the partial-fractions decomposition. That is, if we include other basis 
elements $\tilde{r}_{i \ne n}$ when constructing an ansatz for 
$\tilde{r}_n$, then it will suffice to supplement this set of functions with 
a set that parametrises only the part of $\tilde{r}_n$ not
contained in $\Span(\tilde{r}_{i\neq n})$. 
Naturally, there is a family of such
ans\"atze, with the members corresponding to different choices of terms in the
partial-fractions decomposition to include. Concretely, 
we consider a family of ans\"atze, where each member is of the form
\begin{equation}
  \tilde{r}_n = \frac{\tilde{\mathcal{N}}_{n}}{\prod_j \mathcal{D}_j^{\tilde{q}_{nj}}} + \sum_{i \ne n} c_{ni} \tilde{r}_i,
  \label{eq:ImprovedAnsatz}
\end{equation}
where $\tilde{\mathcal{N}}_{n}$ is an unknown polynomial in spinor brackets
and the $c_{ni}$ are unknown rational numbers. The exponents $\tilde{q}_{nj}$ in
eq.~\eqref{eq:ImprovedAnsatz} are chosen so that the involved denominator
factors are a subset of those in the common denominator form of $\tilde{r}_n$.
Given our one-loop-like motivation, we choose the $\tilde{q}_{nj}$ so that only
one pole of the form $\langle a | b+c |a]^\alpha$ is involved in each member of
the family.
In the example of eq.~\eqref{eq:partial_fractioned_r174}, there are seven
members of the family of ans\"atze, each corresponding to a term in
eq.~\eqref{eq:partial_fractioned_r174}. Taking each member in turn, we sample
the $\tilde{r}_n$ appropriately using finite-field kinematics randomly generated
with
\texttt{lips}\cite{DeLaurentis:2020xar, DeLaurentis:2023qhd}.
We then attempt to fit $\tilde{r}_n$ with each ansatz, until we find that one
is successful, validating our working assumption. If any of the 
working assumptions we made were not valid we would not find a successful 
fit, but in all cases considered in this work we find that our 
assumption holds. 
We note that an important step in this procedure is to parametrise 
polynomials in spinor brackets. This is a non-trivial exercise due to 
momentum conservation and Schouten identities and we make use of the 
systematic solution proposed in ref.~\cite{DeLaurentis:2022otd}.

Having constrained a single $\tilde{r}_n$ with an ansatz of the form given in
eq.~\eqref{eq:ImprovedAnsatz}, we have determined the 
$\tilde{\mathcal{N}}_n$, the $\tilde{q}_{nj}$ and the $c_{ni}$. As we do 
not yet know the analytic form of the $\tilde{r}_{i \ne n}$, we have not 
yet determined the analytic form of $\tilde{r}_n$. Nevertheless, we can 
sidestep this problem by  choosing to change basis and replace 
$\tilde{r}_n$ with the first term in eq.~\eqref{eq:ImprovedAnsatz}. 
In order to fully determine a complete set of linearly independent rational
functions, we then apply this procedure iteratively.
Once this procedure is finished we will need a different matrix $M_{ji}$ 
that relates our basis to the original set of rational functions $r_i$ in
eq.~\eqref{eq:pentagon_function_coefficients_projected_to_basis}.
This matrix can be obtained in a similar way as $M_{ji}$ was determined.

Let us finish by briefly commenting on the benefits of our ansatz procedure. 
We find that the dimension of each ansatz that we make is greatly reduced in
comparison to common denominator form. Indeed, the improved ansatz in
eq.~\eqref{eq:ImprovedAnsatz} involves a numerator of
much lower mass dimension than that in common denominator form, supplemented
only by a few extra parameters for the other rational functions.
In practice we find that the largest ansatz we use has 
only $\mathcal{O}(4000)$ unknowns, which is approximately 10 times smaller
than the largest ansatz in common-denominator form.
Importantly, as we systematically walk through all possible choices of 
ans\"atze we can recycle the numerical evaluations. 
Furthermore, given the small number of unknowns in the improved ansatz, 
the many linear systems that we solve are performed at negligible 
computational cost in comparison to the numerical sampling.
Finally, we note that once a successful small ansatz for an $\tilde{r}_n$ of
the form in eq.~\eqref{eq:ImprovedAnsatz} has been found, we can efficiently
further simplify the analytic form of the $\tilde{r}_n$ by systematically
trying simpler ans\"atze with fewer poles and potentially partial fractioning.

\paragraph{Mandelstam reconstruction}

We also perform the reconstruction computation in Mandelstam
invariants following the approach of ref.~\cite{Abreu:2021asb}. The
pentagon-function coefficients are split into parity-even and
parity-odd parts, and are normalized by the corresponding tree, or a
spinor-weight factor in case the latter vanishes. The
denominators in Mandelstam invariants are obtained with the same
procedure described above, with the difference that a single
univariate slice now suffices since the denominators are little-group
invariant (given by products of alphabet letters). 
We sort the coefficients by ansatz size, and select the
simplest subset which constitutes a basis of the vector space of the
rational functions. The numerator ansatz for the most complicated
parity-even or parity-odd coefficient is now a polynomial of degree 32
in 5 independent variables, corresponding to $58905$ unknown parameters.
The number of evaluations of the remainders required to fit the free
parameters of this ansatz is twice this number, since for each
kinematic point we also require an evaluation at the parity conjugate
point in order to differentiate the parity-even from the party-odd
parts of the pentagon-function coefficients. 
In conclusion, to fit the ansatz in common-denominator form the
number of required evaluations is $117810$ in Mandelstam invariants,
compared to $29059$ in spinor variables. We verified numerically over both
$\mathbb{C}$ and $\mathbb{F}_p$ that the results obtained with the two
computations match.


\section{Results}
\label{sec:results}

\subsection{Efficiency Of Analytic Reconstruction}

In order to show the impact of our reconstruction
procedure on the efficiency of the calculation, we gather here representative
data at various stages of the calculation.
In table~\ref{Tab:SpinorLeastCommonDenominatorInfo}, we display a
summary of the complexity of the amplitudes in common denominator form
when making use of spinor-helicity variables.
Extending the notation of
eq.~\eqref{eq:LittleGroupDefinition}, we
denote the collective little-group weights of $\mathcal{N}_i$ as
$\{\mathcal{N}_i\}$. For comparison, in the
same table, we also reproduce the corresponding information for the
planar finite remainders previously obtained in
ref.~\cite{Abreu:2020cwb}.
In table~\ref{Tab:ActualAnsatzUsed}, we summarize the size of the improved ansatz. 
For each remainder, we provide the mass-dimension and little-group weights of
the most complicated rational function. We use this information to 
construct the ans\"atze with 
the algorithm presented in ref.~\cite{DeLaurentis:2022otd}. The Gr\"obner-basis
computation is performed with \textsc{Singular} \cite{DGPS}. Enumeration of the
spinor monomials is performed with
\textsc{OR-Tools CP-SAT} \cite{ortools}. The resulting dimension of the ansatz
can therefore easily be counted, and we provide this information in the final
column of our tables. We note that this cannot be determined from the mass
dimension alone.

\begin{table}[t]
  \centering
  \begin{tabular}{cccccc}
    \toprule
    \multicolumn{2}{c}{Contribution} & $\text{dim}(\text{span}(r_i))$ & $\text{max}_j([\text{Num}(\tilde{r}_j)]),\,\{\text{Num}(\tilde{r}_{j_{max}})\}$ & \begin{tabular}{@{}c@{}} Common Den. \\ Ansatz Size \end{tabular} \\
    \midrule
    \parbox[t]{2mm}{\multirow{4}{*}{\rotatebox[origin=c]{90}{$\kern-3mm$non-planar}}} & $R^{(2,1)}_{-++}$           &  174 & 48, \{1, -3, -6, 2, 2\} & 29059 \\
    & $R^{(2,\tilde{\NF})}_{-++}$ &  88 & 47, \{4, 4, -5, 3, 4\} & 24582 \\
    & $R^{(2,1)}_{+++}$           &  49 & 21, \{5, 4, 3, 3, 3\} & 1092 \\
    & $R^{(2,\tilde{\NF})}_{+++}$  & 24 & 20, \{2, 4, 6, 6, 6\} & 535 \\
    \midrule
    \parbox[t]{2mm}{\multirow{4}{*}{\rotatebox[origin=c]{90}{$\kern-2mm$planar}}} & $R^{(2,0)}_{-++}$   &  87 & 35, \{-3, 0, 6, -3, -2\} & 7358 \\
    & $R^{(2,\NF)}_{-++}$ &  29& 15, \{-2, -2, 0, -3, -3\} & 378 \\
    & $R^{(2,0)}_{+++}$   &  31 & 20, \{-2, -4, -2, -2, -2\} & 1140 \\
    & $R^{(2,\NF)}_{+++}$ &  6 & 8, \{1, 3, 1, 1, 2\} & 44 \\
    \bottomrule
  \end{tabular}
  \captionsetup{width=.95\linewidth}
  \caption{\label{Tab:SpinorLeastCommonDenominatorInfo}Summary
    of rational function space in common denominator form for the
    non-planar finite remainders (this work), and, for reference, for
    the planar ones (originally calculated in
    ref.~\cite{Abreu:2020cwb}). In the first column, we label the
    associated finite remainder. In the second, we give the number of
    linearly independent rational functions that arise in the
    remainder. In the third, we record the mass dimension and little
    group weights of the most complicated rational function in the
    basis. In the final column, we state the number of terms in the
    common-denominator ansatz for this function.}
\end{table}

\begin{table}[t]
  \centering
  \begin{tabular}{ccccc}
    \toprule
    Contribution & $\text{max}([\tilde{\mathcal{N}}]),\,\{\tilde{\mathcal{N}}\}$ & \begin{tabular}{@{}c@{}} Improved ansatz Size \end{tabular} \\
    \midrule
    $R^{(2,1)}_{-++}$        & 30, \{1, -3, -6, 2, 2\} & 4003 \\
    $R^{(2,\tilde{\NF})}_{-++}$ & 31, \{4, 4, -5, 3, 4\} & 3810 \\
    $R^{(2,1)}_{+++}$        & 21, \{5, 4, 3, 3, 3\} & 1092 \\
    $R^{(2,\tilde{\NF})}_{+++}$ & 20, \{2, 4, 6, 6, 6\} & 535 \\
    \bottomrule
  \end{tabular}
  \caption{\label{Tab:ActualAnsatzUsed}Summary of size of the function
    space of the numerator $\tilde{\mathcal{N}}$  in the improved ansatz
    of eq.~\eqref{eq:ImprovedAnsatz} that was used to reconstruct the analytic
    results. In the second column, we record the mass dimension 
    and little-group
    weights of the most complicated numerator function in the basis. 
    In the last column, we state the number of terms in the numerator ansatz 
    for this function. 
    By comparison to table~\ref{Tab:SpinorLeastCommonDenominatorInfo},
    we see that the improved ansatz has greatly decreased the number of
    required samples. }
\end{table}

In summary, we see that reconstruction of the non-planar finite remainders with
the original common-denominator ansatz requires about four times more
samples than in the planar case.
Furthermore, we see that the sampling requirement for the non-planar
computation is strongly reduced when we use the improved ansatz of
eq.~\eqref{eq:ImprovedAnsatz}. In fact, with $\mathcal{O}(4000)$
required samples, the complexity is below that of the common-denominator ansatz
of the planar amplitudes.
We note that for the all-plus configurations, the ansatz size is
unchanged with respect to table~\ref{Tab:SpinorLeastCommonDenominatorInfo} as
the most complex function does not contain a $\langle c|a+b|c]$ pole and so is
unaffected by the procedure of section~\ref{sec:reconstruction}.
After the analytic reconstruction, we perform further
clean-up following the partial-fraction strategies of
refs.~\cite{DeLaurentis:2019bjh, DeLaurentis:2022otd} and simplify the rational
functions to the point where none has more than about $100$ free coefficients.
The file size of the final results is then dominated by the
matrices of rational numbers, while the rational functions are about one order
of magnitude smaller.
We discuss the ancillary files in more detail in 
section~\ref{sec:AncillaryFiles}.

Let us now briefly comment on the analytic properties of the amplitudes.
Firstly, we verified that all iterated integrals with the letter $W_{31} =
\tr_5$ cancel out in all finite remainders.
Secondly, we find that only 162 independent combinations of irreducible weight-four functions are required to express
$\mathcal{H}^{(2)}$, a number that is significantly lower than the
472 required to span the full space of irreducible weight-four pentagon functions \cite{Chicherin:2020oor}.
It is also interesting to compare the functions arising in the planar triphoton
amplitudes with those for leading-color three-jet production. We find that only
48 independent combinations of irreducible weight-four functions are required to
express the planar triphoton amplitudes and that they span a proper subspace of
those needed for three-jet production.

\subsection{Validation}

In order to validate our computation, we have performed various checks at
intermediate stages, as well as on the final result which we now discuss.
To verify the surface terms, we produced numerical reduction tables on one
phase-space point using \texttt{FIRE6}~\cite{Smirnov:2019qkx} and checked that
the surface terms reduce to zero.
The evaluation of the amplitudes with the numerical unitarity method was
carried out by the well-tested code \textsc{Caravel}~\cite{Abreu:2020xvt}.
The construction of finite remainders was also carried out within
\textsc{Caravel}, thus confirming the expected pole-structure on every
phase-space point. In particular we confirm that the contribution
$A^{(2,\NFqq)}$ is finite, which is expected as its coupling structure is non-zero for the first time at two loops.
This furthermore implies that this contribution is invariant under scale variations, which we also confirm.
To validate the reconstruction methodology, we confirm that the analytic results
reconstructed in spinor variables agree with the results obtained
using established Mandelstam-variable reconstruction techniques. Furthermore, 
we check that the analytic results match numerical evaluations from 
\textsc{Caravel} on finite fields of different characteristic.
For the planar contributions $R^{(2,0)}$ and $R^{(2,\NF)}$, we find full
agreement between our result and a previous calculation \cite{Abreu:2020cwb}.
Finally, we evaluated the remainders in a number of collinear
configurations to verify the expected singular behavior. In particular, we
validate the sign of the $R^{(2,\NFqq)}$ contribution by numerically
checking that its behavior in collinear limits is consistent with the universal
expectation. To this end, we make use of the known two-loop four-point
amplitudes of ref.~\cite{Glover:2003cm}.

\subsection{Ancillary Files}
\label{sec:AncillaryFiles}

In the ancillary files associated to this paper, we present all finite
remainders through two loops. All finite remainders are presented in the form
\begin{equation}
  R = \tilde r_j M_{ji} h_i \, \label{eq:form_of_finite_remainder},
\end{equation}
where $\tilde{r}_j$ are rational functions, $M_{ji}$ is a matrix of rational
numbers and the $h_i$ are monomials of pentagon functions.
The ancillary files are provided in \texttt{Mathematica} format, organized by
gauge-invariant contributions, with naming conventions based on couplings and
helicities.

To aid with the understanding of the ancillary files, 
we provide assembly scripts that
evaluate the results of this paper and reproduce the benchmark values in
appendix~\ref{sec:reference-values}, which are also found in the file
\texttt{targets.m}. For ease of use, we provide these assembly scripts both in
\texttt{Mathematica} (\texttt{amp\_eval.m}) and in \texttt{Python}
(\texttt{amp\_eval.py}) format. 
The \texttt{Python} tests can be run with \texttt{pytest} \cite{pytest7.1}.
The assembly scripts make use of a series of raw analytic files for each
amplitude. Each finite remainder is associated to a subfolder of
\texttt{anc/amplitudes} which contains the following files:
\begin{itemize}
\item \texttt{BasisCoefficients.m}: The rational functions $\tilde{r}_j$ 
in eq.~\eqref{eq:form_of_finite_remainder}. Each element of the
list is either a rational function of spinor variables or a
list of integers. A list of integers represents a permutation of the spinors in
the previous element in the list. For a list of integers \texttt{\{i1, i2,
i3, i4, i5\}} the permutation to be applied is
  \begin{equation*}
    (i_1 i_2 i_3 i_4 i_5 \rightarrow 12345).
  \end{equation*}
\item \texttt{Matrix.m}: The matrix of rational numbers $M_{ji}$ in eq.~\eqref{eq:form_of_finite_remainder}. The matrices are given
  in the sparse coordinate list (COO) format. In this format,
  only the non-zero values are stated and all unspecified entries are
  implicitly zero. The non-zero entries in COO format are specified as
  \texttt{\{row index, column index\} -> value}. 
\item \texttt{BasisPentagons.m}: 
The pentagon functions in eq.~\eqref{eq:form_of_finite_remainder}.
\end{itemize}
Cached values for the pentagon-function monomials at the kinematic
point in \cref{sec:reference-values} 
(see eq.~\eqref{eq:benchmarkphasespacepoint}) are given in
\texttt{FValues.m}. The files \texttt{Benchmark.m} contain target
values for the individual pentagon-function coefficients $r_i =
\tilde{r}_j M_{ji}$ at this kinematic point.

\paragraph[Explicit example: $R^{(2,\NFqq)}_{+++}$]{\boldmath Explicit example: $R^{(2,\NFqq)}_{+++}$}

In order to elucidate the structure of the results accompanying this
paper, we discuss in detail the structure of one of the non-planar
finite remainders, $R^{(2,\NFqq)}_{+++}$.
The dimension of the rational function space is 24 
(see \cref{Tab:SpinorLeastCommonDenominatorInfo}).  
The number of contributing pentagon-function monomials is 71.
The sparse rational matrix $M$ has dimension $24 \times 71$, with
177 nonzero entries.
The basis of rational functions
$\tilde{r}$ can be written as
\begin{align}
  \begin{split}
    \tilde{r} =
    \Big\{&\tilde{r}_{1} \,,\; \tilde{r}_{1}\big|_{453 \rightarrow 345} \,,\; \tilde{r}_{1}\big|_{534 \rightarrow 345} \,,\; \tilde{r}_{4} \,,\;  \tilde{r}_{4}\big|_{453 \rightarrow 345} \,,\; \tilde{r}_{4}\big|_{534 \rightarrow 345} \,,\; \tilde{r}_{4}\big|_{543 \rightarrow 345} \,,\; \tilde{r}_{4}\big|_{435 \rightarrow 345} \,,\; \\
    &\,\, \tilde{r}_{4}\big|_{354 \rightarrow 345} \,,\; \tilde{r}_{10} \,,\; \tilde{r}_{10}\big|_{453 \rightarrow 345} \,,\; \tilde{r}_{10}\big|_{534 \rightarrow 345} \,,\; \tilde{r}_{13} \,,\; \tilde{r}_{13}\big|_{453 \rightarrow 345} \,,\; \tilde{r}_{13}\big|_{534 \rightarrow 345} \,,\; \tilde{r}_{16} \,,\;  \\
    &\,\, \tilde{r}_{16}\big|_{453 \rightarrow 345} \,,\; \tilde{r}_{18} \,,\; \tilde{r}_{18}\big|_{453 \rightarrow 345} \,,\; \tilde{r}_{18}\big|_{534 \rightarrow 345} \,,\; \tilde{r}_{18}\big|_{543 \rightarrow 345} \,,\; \tilde{r}_{18}\big|_{435 \rightarrow 345} \,,\; \tilde{r}_{23} \,,\; \tilde{r}_{24} \Big\} \, ,
  \end{split}
\end{align}
where we make explicit that many of the basis functions are obtained by
permuting momenta in previous elements.
It is interesting that owing to
the symmetries of the rational function space it suffices to give
8 explicit spinor-helicity expressions.
The basis elements
that generate the others under permutations are given by
\begin{align}
  \begin{split}
    \tilde{r}_1 &= 16\frac{[13]^2}{[12]⟨45⟩^2} \, , \\[1.5mm]
    \tilde{r}_4 &= -8\frac{⟨12⟩[13][34](⟨34⟩⟨12⟩+2⟨24⟩⟨13⟩)}{⟨14⟩^2[14]⟨15⟩⟨25⟩⟨34⟩} \, , \\[1.5mm]
    \tilde{r}_{10}^D &= -24\frac{⟨12⟩⟨24⟩[34]}{⟨14⟩⟨15⟩⟨25⟩⟨34⟩} \, , \quad
    \tilde{r}_{10}^S = -16\frac{⟨12⟩[15][34]}{⟨15⟩⟨34⟩⟨1|2+5|1]} \, , \\[1.5mm]
    \tilde{r}_{10} &= \tilde{r}_{10}^D + \tilde{r}_{10}^D\big|_{345 \rightarrow 435} + \tilde{r}_{10}^S \, , \\[1.5mm]
    \tilde{r}_{13}^D &= 24\frac{⟨12⟩[15]}{⟨13⟩⟨34⟩⟨45⟩} -
    48\frac{[15]⟨25⟩^2}{⟨24⟩⟨35⟩^2⟨45⟩} +
    24\frac{⟨12⟩^2[15]}{⟨13⟩⟨14⟩⟨23⟩⟨45⟩} \, , \\[1.5mm]
    \tilde{r}_{13}^S &=  
    48\frac{[15]⟨24⟩}{⟨34⟩^2⟨45⟩} -
    24\frac{[15]⟨25⟩}{⟨34⟩⟨35⟩⟨45⟩} \, , \\[1.5mm]
    \tilde{r}_{13} &= \tilde{r}_{13}^D + \tilde{r}_{13}^D\big|_{345 \rightarrow 435} + \tilde{r}_{13}^S \, , \\[1.5mm]
    \tilde{r}_{16}^D &= - 48\frac{⟨23⟩^2[23]}{⟨15⟩⟨34⟩^2⟨35⟩} +
    24\frac{[23]⟨24⟩⟨25⟩}{⟨15⟩⟨34⟩⟨45⟩^2} \, , \\[1.5mm]
    \tilde{r}_{16}^S &= 16\frac{⟨12⟩[13][45]}{⟨13⟩⟨45⟩⟨1|2+3|1]} -
    48\frac{⟨12⟩⟨23⟩[23]}{⟨14⟩⟨15⟩⟨34⟩⟨35⟩} \, , \\[1.5mm]
    \tilde{r}_{16} &= \tilde{r}_{16}^D + \tilde{r}^D_{16}\big|_{345 \rightarrow 354} + \tilde{r}_{16}^S \, ,
  \end{split}
\end{align}
where the superscripts refer to doublets (D) and singlets (S) under the
permutation transformation used to express the coefficient. We omit the three functions
$\tilde{r}_{18}, \tilde{r}_{23}$ and $\tilde{r}_{24}$ as they are too
complicated to print in the text. 

\subsection{Numerical Evaluation}
\label{sec:numerical-evaluation}

To facilitate the use of our results in phenomenological applications, we have
implemented our analytic expressions into the \texttt{C++} library
\texttt{FivePointAmplitudes} \cite{FivePointAmplitudes}, making use of
\texttt{PentagonFunctions++} \cite{Chicherin:2020oor} for the 
numerical evaluation of the pentagon functions.
For use in such applications, it is important that our implementation is
capable of producing numerically-stable results while maintaining reasonable
evaluation times.
To demonstrate the numerical performance of our implementation, we study the
(helicity) finite remainders and the hard function $\mathcal{H}$ on a sample of
$100k$ phase-space points generated by
\texttt{MATRIX~v2}~\cite{Grazzini:2017mhc}. We adopt the phase-space definition
from ref.~\cite{Kallweit:2020gcp}, and we set the renormalization scale to $\mu
= m_{\gamma\gamma\gamma}$.

\begin{figure}[ht!]
  \centering

  \begin{subfigure}[t]{0.7\textwidth}
    \includegraphics[width=1\textwidth]{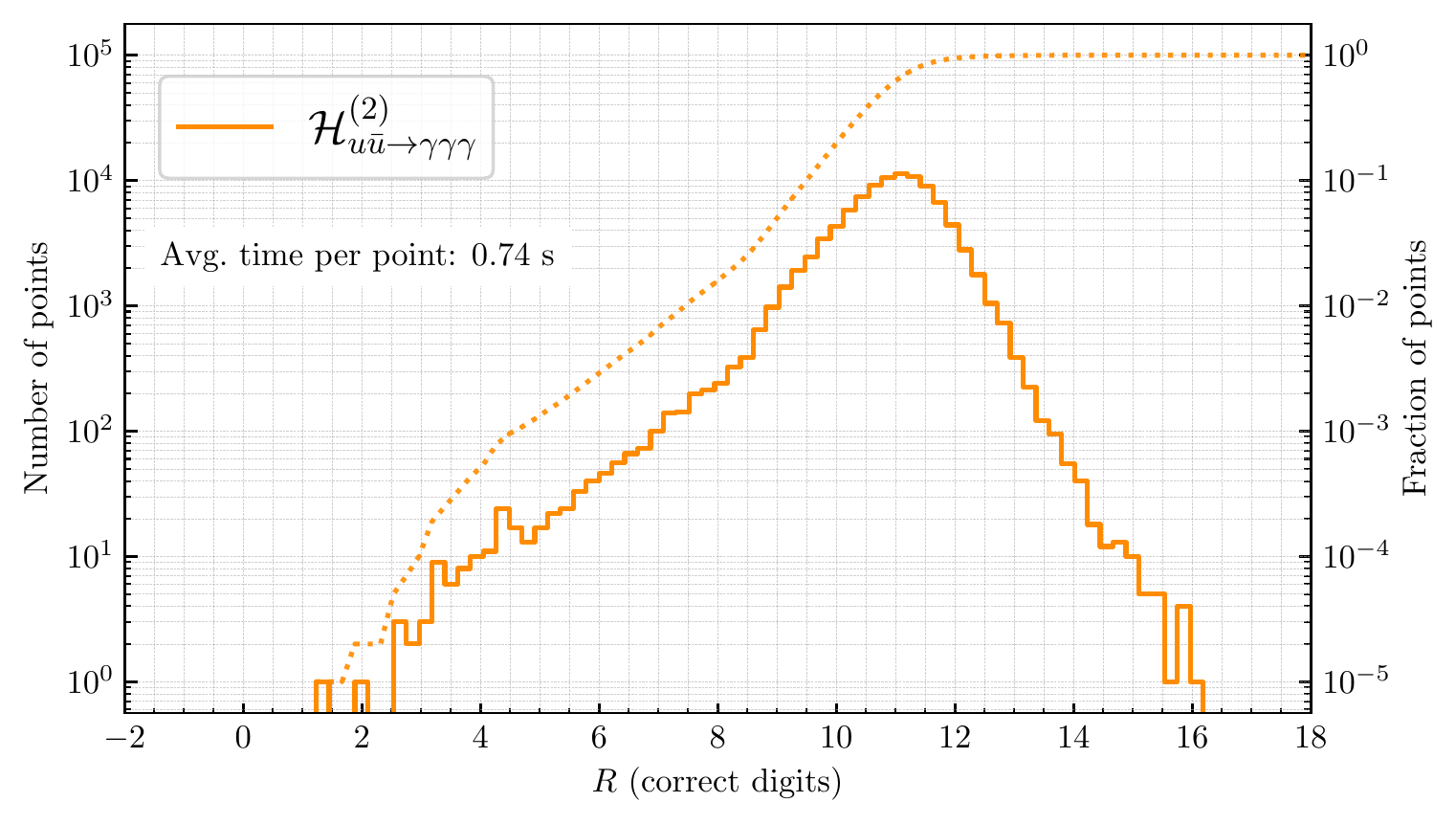}
    \caption{Distributions of correct digits for full $\mathcal{H}^{(2)}$ with
      five active flavours, i.e.\ $\NC=3,$ $\NF=5$. The dashed line shows the
      cumulative distribution.}
    \label{fig:H2stability-full}
  \end{subfigure}

  \vspace{8mm}
  
  \begin{subfigure}[t]{0.7\textwidth}
    \includegraphics[width=1\textwidth]{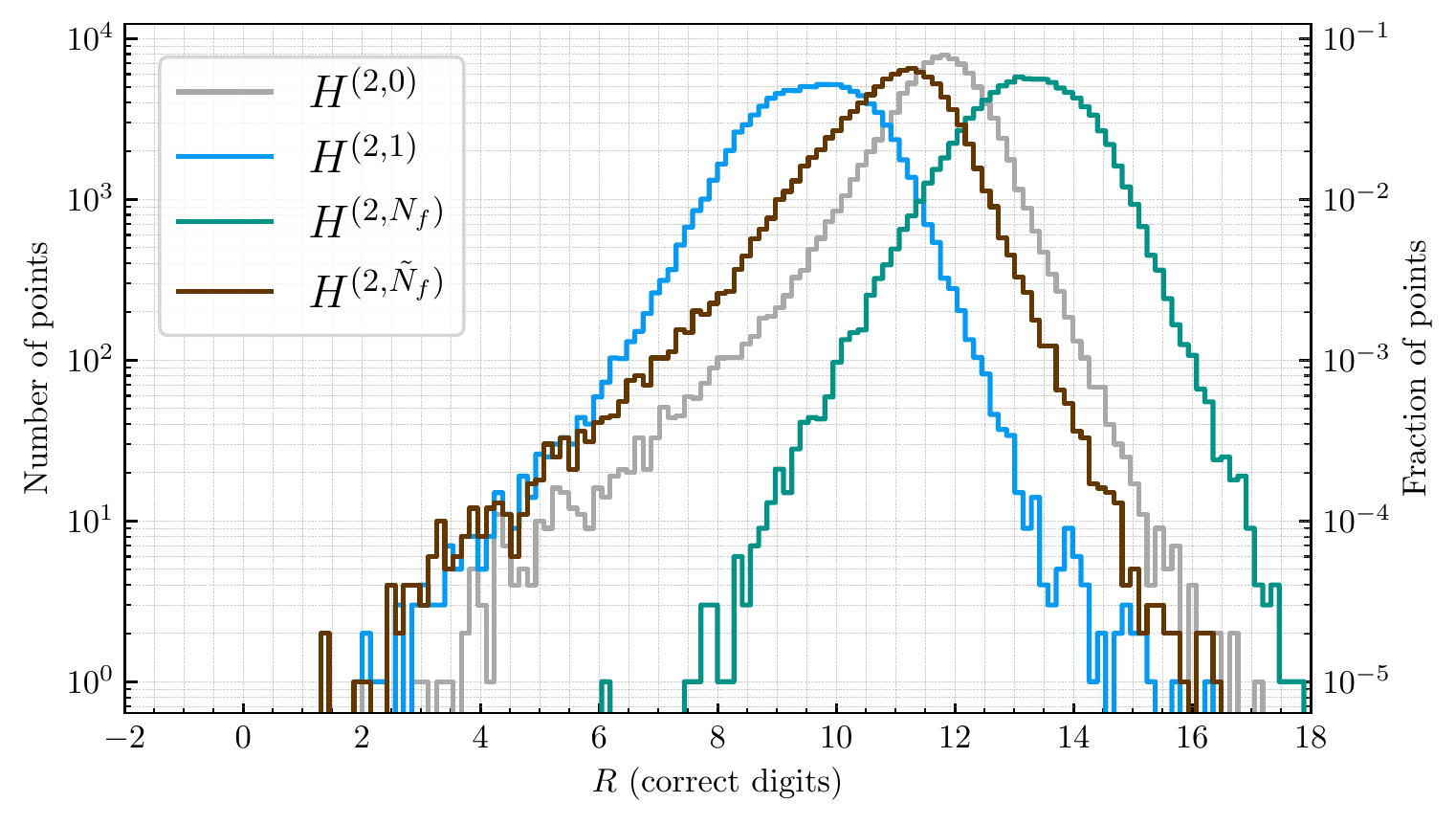}

    \caption{Relative error of the contributions defined in eq.~\eqref{eq:hard-function} to the hard function $\mathcal{H}^{(2)}$.}
    \label{fig:H2stability-separate}
  \end{subfigure}
  
  \vspace{4mm}

  \caption{Distributions of correct digits $R$ (see
    eq.~\eqref{eq:correct-digits}) that characterize the numerical performance
    of the \texttt{C++} implementation of our analytic results in
    \texttt{FivePointAmplitudes} \cite{FivePointAmplitudes}. The phase-space
    sample of $100k$ phase-space points is defined in
    ref.~\cite{Kallweit:2020gcp} and generated by \texttt{MATRIX v2}~\cite{Grazzini:2017mhc}.
    The renormalization scale is set to $\mu = m_{\gamma\gamma\gamma}$. 
  }
  \label{fig:numerical-stability}

\end{figure}

We characterize the numerical stability of our implementation in
\cref{fig:numerical-stability}, where we plot the distributions of correct
digits $R$,
\begin{equation} \label{eq:correct-digits}
  R \coloneqq -\log_{10}\abs{ 1 - \frac{X_\text{double}}{X_\text{quad} }}\,,
\end{equation}
for various quantities $X$. 
In this way, we use a quadruple-precision evaluation ($X_\text{quad}$) to calculate the accuracy of the double-precision evaluation ($X_\text{double}$) on each point.
To catch and correct unstable evaluations we recycle the precision rescue system developed in ref.~\cite{Abreu:2021oya} for the amplitudes
describing three-jet production at hadron colliders.
We note however that in this case it triggers only on a few points from the whole sample, so its effect on the evaluation time is insignificant.
In \cref{fig:H2stability-full} we show the $R$-distribution for the hard function $\mathcal{H}$ defined in \cref{eq:hard-function}
for the dominant partonic process $u\bar{u}\to \gamma\gamma\gamma$ with five active
massless flavors, i.e.\ $\NC=3,$ $\NF=5$.
We observe overall excellent numerical stability, and the average evaluation time of less than a second on a single CPU core.
We can therefore conclude that our implementation is suitable for phenomenological applications. 
Since the analytic complexity of the subleading-color contributions is notably higher than that of the leading contributions, 
it is of interest to compare their relative numerical stability.
In \cref{fig:H2stability-separate} we show the $R$-distributions for each of the four contributions to $\mathcal{H}$ separately. 
Indeed, we observe that the subleading functions
$H^{(2,1)}$ and $H^{(2,\NFqq)}$ are slightly less numerically stable than the leading functions $H^{(2,0)}$, $H^{(2,\NF)}$. 
Nevertheless, their numerical behavior is clearly adequate for the anticipated phenomenological applications.

\begin{figure}[htb]
  \centering
  \includegraphics[width=0.7\textwidth]{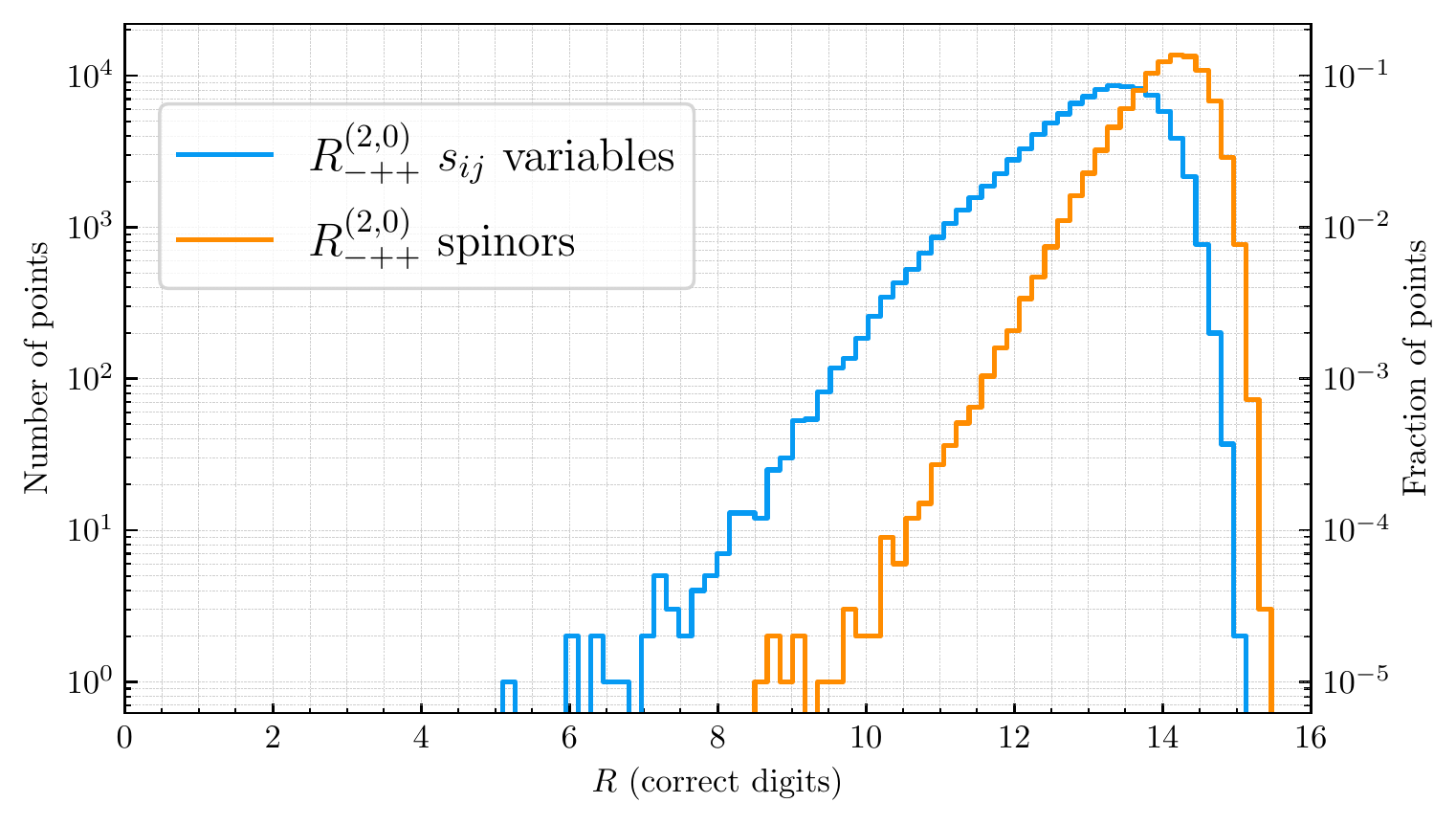}
  \caption{
    Numerical stability of the rational functions contributing to the finite remainder $R^{(2,0)}_{-++}$ in two different representations: 
    one in terms of Mandelstam invariants and the other in terms of
spinor-helicity variables. At each phase-space point, we plot the lowest number of
correct digits
$R$ over the pentagon-function coefficients $r_i$ of
\cref{eq:finite-remainder-as-weighted-sum}. See the caption of
\cref{fig:numerical-stability} for details of the phase-space definition.
  }
  \label{fig:Rmpp-planar-spinor-sij}
\end{figure}

Finally, let us recall that we employ a spinor representation of
rational coefficients in our numerical implementation. Nevertheless, as
discussed in~\cref{sec:reconstruction}, we also reconstructed the rational
coefficients in terms of Mandelstam variables. With this data at hand,
it is interesting to study the impact of the choice of variables 
on the numerical stability of the analytic expressions.
In \cref{fig:Rmpp-planar-spinor-sij} we show the $R$-distributions for the
coefficients $r_i$ of pentagon functions as defined in
\cref{eq:finite-remainder-as-weighted-sum}.
Specifically, on each phase-space point we compute the value of $R$ for each
$r_i$ and plot $\text{min}_{r_i}(R)$. As a representative example, 
we show this distribution for the finite remainder $R_{-++}^{(2,0)}$, first
using the compact spinor representation of the ancillary files, and then in a
representation in terms of Mandelstam variables that has been put into a partial
fractions representation using MultivariateApart~\cite{Heller:2021qkz}.
We see that the tail of the distribution of $R$ in the spinor case is improved
by about two digits with respect to the Mandelstam case, a behavior that we find is
consistent across all helicity amplitudes.
It is therefore evident that the spinor representation improves the numerical stability
compared to the representation through Mandelstam invariants. 
In spite of this, for the contribution $H^{(2,0)}$ we observe overall
similar behavior as in ref.\ \cite{Abreu:2020cwb}. This implies that the
numerical accuracy of $\mathcal{H}^{(2)}$ is limited by the accuracy of the
pentagon-function evaluation.


\section{Subleading Color Corrections To Hard Function}
\label{sec:H-functions-effect}

In all phenomenological studies of triphoton production at hadron colliders
to date, the double-virtual NNLO QCD corrections have been included in the
leading-color approximation.
Given our results for the subleading color contributions to the double-virtual
corrections, it is interesting to consider the numerical impact of these
corrections.
Specifically, in refs.~\cite{Chawdhry:2019bji,Kallweit:2020gcp} the hard function was taken
to be
\begin{equation} \label{eq:h2-LC}
  \mathcal{H}^{(2)} \; \to \; \mathcal{H}^{(2)}_{\text{l.c.}} \coloneqq \frac{\NC^2}{4} H^{(2,0)} \, ,
\end{equation}
instead of the full result in \cref{eq:hard-function}.
In this section, we study the impact of the subleading-color contributions
that we have calculated in this work relative to
$\mathcal{H}^{(2)}_{\text{l.c.}}$. We focus on the dominant partonic
channel $u\bar{u} \to \gamma\gamma\gamma$, and we consider
distributions of the relative sizes of corrections with respect to the leading-color result,
\begin{equation}
  \delta \mathcal{H}_x^{(2)} = \frac{\Delta \mathcal{H}_x^{(2)}}{\mathcal{H}^{(2)}_{\text{l.c.}}}
  \, ,
\end{equation}
over the same phase-space points used in
\cref{sec:numerical-evaluation}. Here $\Delta \mathcal{H}_x^{(2)}$ is
one of the three corrections in \cref{eq:hard-function}:
\begin{align}\begin{split}\label{eq:subleading-color-corrections}
    \Delta \mathcal{H}_{N_c}^{(2)} \, & \coloneqq \, - \frac{1}{4}(H^{(2,0)}+H^{(2,1)})  + \frac{1}{4 \NC^2} H^{(2,1)} \, , \\
    \Delta \mathcal{H}_{N_f}^{(2)} \, & \coloneqq \, C_F T_F \NF H^{(2,\NF)} \, , \\
    \Delta \mathcal{H}_{Q_f^2}^{(2)} \, & \coloneqq \, C_F T_F \left(\sum_{f=1}^{\NF} Q_f^2 \right) \, H^{(2,\NFqq)} \, ,
\end{split}\end{align}
such that $\mathcal{H}^{(2)} = \mathcal{H}^{(2)}_{\text{l.c.}} + \Delta \mathcal{H}_{N_c}^{(2)} + \Delta \mathcal{H}_{N_f}^{(2)} + \Delta \mathcal{H}_{Q_f^2}^{(2)}$,
and $\delta{\mathcal{H}}^{(2)} \coloneqq  \mathcal{H}^{(2)}/\mathcal{H}^{(2)}_{\text{l.c.}}-1$.
We note that the $\delta \mathcal{H}_x$ are scheme dependent, and can vary
substantially between different schemes.
In \cref{fig:correction-sizes}, we plot the distribution of the $\delta
\mathcal{H}_x$ over phase space in the $q_T$, $\MSbar$ and Catani schemes (see
\cref{sec:IR-scheme-change} for our scheme definitions).
The average correction size in \cref{fig:correction-sizes} should provide a
reasonable proxy for the subleading-color effects in fiducial cross sections,
whereas the shape demonstrates how the corrections vary over phase space.
Our phase-space sample and conventions are the same as that used in the studies
of numerical stability. Specifically, our 100k sample points are taken from the phase
space used in the Monte-Carlo cross-section computation of
ref.~\cite{Kallweit:2020gcp}, as generated by \texttt{MATRIX
  v2}~\cite{Grazzini:2017mhc}. We work with five active flavors, 
i.e.\ $N_f = 5$, and set $N_c = 3$ and  the renormalization scale to 
$\mu = m_{\gamma \gamma \gamma}$.

\begin{figure}[htp!]
  \centering
  \begin{subfigure}[c]{0.9\textwidth} %
    \centering
    \includegraphics[width=1.0\textwidth]{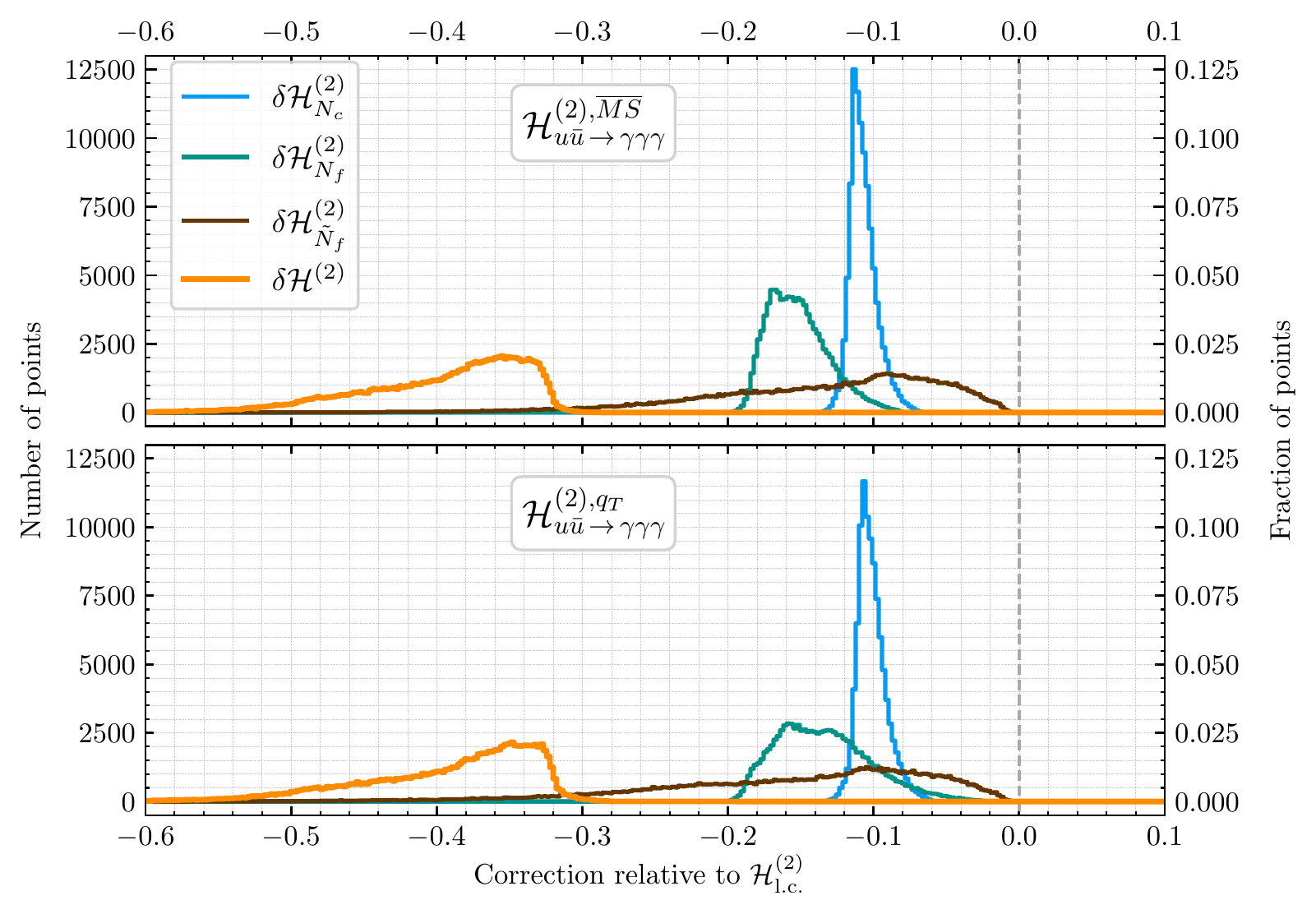}
    \caption{Distribution of subleading-color corrections in the $q_T$ and $\MSbar$ schemes.}
    \label{fig:correction-sizes-qT}
  \end{subfigure}

  \vspace{10mm}
  
  \begin{subfigure}[c]{0.9\textwidth} %
    \centering
    \includegraphics[width=1.0\textwidth]{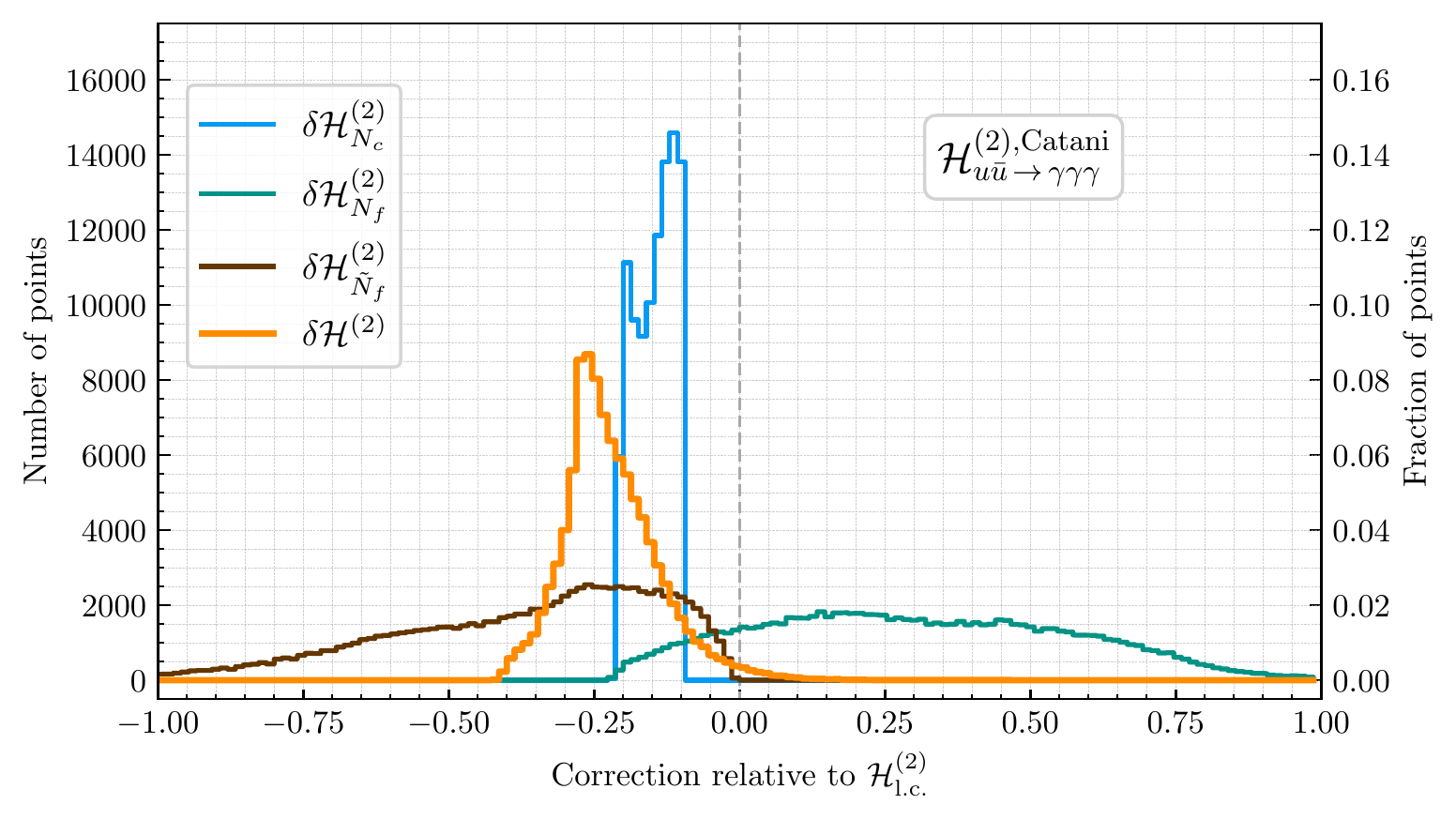}
    \caption{Distribution of subleading-color corrections in the Catani scheme.}
    \label{fig:correction-sizes-catani}
  \end{subfigure}

  \vspace{5mm}
  
  \caption{Distribution of separate subleading-color corrections
          $\delta \mathcal{H}_x$ defined in \cref{eq:subleading-color-corrections} to the
          the hard function $\mathcal{H}$ alongside the total correction $\delta
          \mathcal{H}$. The phase-space sample of $100k$ phase-space points is defined in
          ref.~\cite{Kallweit:2020gcp} and generated by \texttt{MATRIX
          v2}~\cite{Grazzini:2017mhc}. The renormalization scale is set to $\mu =
          m_{\gamma\gamma\gamma}$.}
  \label{fig:correction-sizes}
\end{figure}

Let us now discuss key features of these distributions.
First, we consider the impact of the choice of IR subtraction scheme and see in
\cref{fig:correction-sizes} that the $q_T$ and $\MSbar$ schemes behave
similarly, and both differ noticeably from the Catani scheme.
Next, we note that a highly peaked distribution implies a simple correction by a
factor in all differential cross sections, while a smeared distribution implies
that the corrections are observable-dependent.
Focusing specifically on \cref{fig:correction-sizes-qT} (the schemes used
in refs.~\cite{Chawdhry:2019bji,Kallweit:2020gcp}), we see that all corrections
are negative, but behave differently over phase space.
The correction from the $\NC$-suppressed terms is sharply peaked at about $-10\%$, in good agreement with the common subleading-color behavior. 
The $H^{(2,\NFqq)}$ corrections on the contrary demonstrate significant phase-space dependence.
On average, the combined correction is about $-35\%$, and the shape of the distribution suggests that for some observables the correction could reach up to about $-50\%$.
Nevertheless, taking into account the smallness of the overall contribution of $\mathcal{H}^{(2)}$ to NNLO cross sections,
we expect that the approximation employed in 
refs.~\cite{Chawdhry:2019bji,Kallweit:2020gcp} should be largely adequate. 
Finally, it is worth noting that in the $d\bar{d}$ partonic channel the
contribution from $H^{(2,\NFqq)}$ to $\mathcal{H}^{(2)}$ is four times larger
than in $u\bar{u}$, while all other contributions are unchanged. Therefore,
$H^{(2,\NFqq)}$ becomes numerically as important as
$\mathcal{H}^{(2)}_\text{l.c.}$ in this channel and the approximation of
\cref{eq:h2-LC} is no longer valid.
However, the overall contribution of the $d\bar{d}$ channel to the cross section is highly suppressed by the ratio of quark electric charges, so this effect is invisible.

\FloatBarrier

\section{Conclusion}
\label{sec:conclusion}

In this work, we have computed the complete set of two-loop helicity amplitudes that contribute to
the NNLO QCD corrections for triphoton production at hadron colliders, 
including all previously unknown non-planar contributions.
We have derived compact analytic expressions for the two-loop finite remainders 
that we make available as supplementary material.
In addition, we have implemented our analytic results in a C++
library, providing a stable and efficient numerical
evaluation that is suitable for phenomenological applications.

To carry out our computation, we employed the numerical unitarity approach as
implemented in \textsc{Caravel}, which we have suitably extended to take care of
the reduction of non-planar five-point integral topologies.
To handle the ensuing increase in complexity, we have proposed a new approach
for the construction of surface terms that capitalizes on formulating 
the associated syzygy problem in embedding space, 
and on the numerical nature of the computational framework.
To reconstruct the analytic form of the results from numerical evaluations 
of the amplitude over finite fields we have constructed an optimized ansatz 
in spinor-helicity variables.
This resulted in a more efficient reconstruction procedure, with the
most complicated ansatz having only around 4000 free parameters.
We anticipate that the advancements in both of these technical aspects will 
greatly assist in computations of two-loop amplitudes with more 
kinematic scales  that currently pose significant computational challenges.

Using our results, we have estimated the effect of including the subleading-color corrections in the double-virtual
contributions to the cross section. We have found that, on average, 
over a representative sample of phase space,
the two-loop hard function receives an extra negative contribution of
around 25-35\%. The exact effect is observable and scheme dependent.
Due to an overall small contribution from double-virtual corrections to 
the NNLO cross sections \cite{Chawdhry:2019bji,Kallweit:2020gcp},
we conclude that the approximation employed in
ref.~\cite{Chawdhry:2019bji,Kallweit:2020gcp,ATLAS:2021lsc} should be valid within a few percent, 
although in practice it is important to verify this conclusion 
for each observable.
Nevertheless, it is clear that in view of the excellent numerical stability and
efficiency of our results, including the subleading-color corrections to the
double-virtual contributions in future phenomenological studies
of triphoton production is straightforward.

\section*{Acknowledgements}

VS gratefully acknowledges the computing resources provided by the Max Planck Institute for Physics. 

\paragraph{Funding information}
The work of BP was supported by the European Union’s Horizon 2020 research
and innovation program under the Marie Sklodowska-Curie grant agreement No.896690 (LoopAnsatz).
MK’s work was funded by the German Research Foundation (DFG) within the Research
Training Group GRK 2044. 
VS has received funding from the European Research Council (ERC) under the European 
Union's Horizon 2020 research and innovation programme grant agreement 101019620 (ERC Advanced Grant TOPUP).


\begin{appendix}

\section{IR Renormalization}
\label{sec:UV-IR}

\subsection{Conventions In Catani Scheme}\label{sec:catani}

The remainders $\mathcal{R}$ are defined in \cref{eq:remainder2l}
using the functions $\mathbf{I}^{(1)}$ and $\mathbf{I}^{(2)}$
\cite{Anastasiou:2002zn,Glover:2003cm} which we quote here for
convenience,
\begin{align}
  \begin{split}\label{eq:catanii}
    \mathbf{I}^{(1)} &(\epsilon)  = C_F \frac{e^{\gamma_E\epsilon}}{\Gamma(1-\epsilon)}\left(\frac{1}{\epsilon^2}+\frac{3}{2\epsilon}\right)\left(-\frac{s_{12}}{\mu^2}-\ii 0\right)^{-\epsilon}, \\
    \mathbf{I}^{(2)} &(\epsilon) = 
    \frac{1}{2}\mathbf{I}^{(1)}(\epsilon)\mathbf{I}^{(1)}(\epsilon)-\frac{\beta_0}{\epsilon}\mathbf{I}^{(1)}(\epsilon)+\frac{e^{-\gamma_E\epsilon}\Gamma(1-2\epsilon)}{\Gamma(1-\epsilon)}\left(\frac{\beta_0}{\epsilon}+K\right)\mathbf{I}^{(1)}(2\epsilon)-\mathbf{H}(\epsilon).
  \end{split}
\end{align}
In $\mathbf{I}^{(2)}(\epsilon)$, we have introduced the functions
\begin{align}
  \begin{split}
    K & =\left(\frac{67}{18}-\frac{\pi^2}{6}\right) C_A - \frac{10}{9} T_F N_F, \qquad
    \mathbf{H}(\epsilon)=\frac{e^{\gamma_E\epsilon}}{2\epsilon\Gamma(1-\epsilon)} H_q, \\ 
    H_q  &= \left( \frac{\pi^2}{2} -6 \zeta_3 -\frac{3}{8}\right) C_F^2 + \left(\frac{13}{2} \zeta_3 + \frac{245}{216} - \frac{23}{48}\pi^2\right) C_A C_F +  \left( \frac{\pi^2}{12}-\frac{25}{54} \right) T_F C_F \NF.
  \end{split}
\end{align}

\subsection{IR-Subtraction Scheme Change}
\label{sec:IR-scheme-change}

Consider two different IR subtraction schemes, in which finite remainders are
\begin{equation} \label{eq:remainderDef}
  \mathcal{R} = \mathbf{I} \mathcal{A}, \qquad \tilde{\mathcal{R}} = \tilde{\mathbf{I}} \mathcal{A}\,,
\end{equation}
where we suppress the helicity labels.
We write the difference between the squared finite remainders as 
\begin{equation}
  \Delta \coloneqq \abs*{\tilde{\mathcal{R}}}^2 - \abs*{\mathcal{R}}^2 = \left(\abs*{\tilde{\mathbf{I}}}^2 - \abs*{\mathbf{I}}^2\right) \abs*{\mathcal{A}}^2\,,
\end{equation}
where the absolute values do not include helicity sums. Overall
colour factors have been removed as in \cref{eq:def_A} 
and no colour sums are implied.

Upon expansion to the second order in $\alpha_s$ we find
  \begin{align}\begin{split}
    \Delta^{(1)} & = \delta^{(1)} \abs*{\mathcal{A}^{(0)}}^2,  \qquad \delta^{(1)} \coloneqq  2 \Re{\tilde{\mathbf{I}}^{(1)} - \mathbf{I}^{(1)}}\,, \\
    \Delta^{(2)} & = \bar\delta^{(2)} \abs*{\mathcal{A}^{(0)}}^2  + \delta^{(1)} 2 \Re{ \mathcal{A}^{(0)\star} \mathcal{A}^{(1)}}, 
      \quad \bar\delta^{(2)} \coloneqq 2 \Re{\tilde{\mathbf{I}}^{(2)} - \mathbf{I}^{(2)}} -  \abs*{\tilde{\mathbf{I}}^{(1)}}^2 + \abs*{\mathbf{I}^{(1)}}^2\,.
\end{split}\end{align}
We can rewrite $\Delta^{(2)}$ such that it is defined explicitly through finite quantities as 
\begin{equation}
  \Delta^{(2)}  =  \delta^{(2)}\abs*{\mathcal{A}^{(0)}}^2 + \delta^{(1)} 2 \Re{\mathcal{A}^{(0)\star} \mathcal{R}^{(1)}}, \qquad \delta^{(2)} \coloneqq  \bar\delta^{(2)}  + \delta^{(1)} 2 \Re{\mathbf{I}^{(1)}}\,.
\end{equation}

Using the definitions in \cref{eq:hard-function-def}, and taking advantage of
the factorization of $\abs*{\mathbf{I}}^2$ in the sum, we can therefore write
the scheme shift for the hard functions as
\begin{align}\begin{split}
    \tilde{\mathcal{H}}^{(1)} &= \mathcal{H}^{(1)}  + \delta^{(1)}, \\
    \tilde{\mathcal{H}}^{(2)} &=  \mathcal{H}^{(2)} + \delta^{(1)} \mathcal{H}^{(1)} + \delta^{(2)}.
\end{split}\end{align}
dependence of the remainder definition in 
\cref{eq:remainderDef} enters in a linear way, 
the conversion formulas apply equally to helicity-summed hard 
functions $\mathcal{H}^{(L)}$.

Let us consider converting the finite remainders in the Catani scheme, that is employed in this work, to the ones defined in the $\MSbar$ scheme. 
Taking the definitions of $\tilde{\mathbf{I}} = \mathbf{I}^{\MSbar}$ from 
ref.\ \cite{Chawdhry:2020for},%
\footnote{Explicitly, $\tilde{\mathbf{I}}^{(1)} = - \mathbf{Z}^{(1)}/2$,  
$\tilde{\mathbf{I}}^{(2)} = ((\mathbf{Z}^{(1)})^2- \mathbf{Z}^{(2)})/4$,
with the $\mathbf{Z}^{(i)}$ defined in ref.\ \cite{Chawdhry:2020for}.}
we obtain
\begin{align}
    \delta^{(1)}_{\MSbar} &= C_F \left(\frac{7 \pi ^2}{6}+3 l_{\mu } - l_{\mu }^2\right)\,, \notag\\
    \delta^{(2)}_{\MSbar} &= C_F^2 \left(\frac{49 \pi ^4}{72}+\frac{7 \pi ^2}{2} l_{\mu }+\left(\frac{9}{2}-\frac{7 \pi ^2}{6}\right) l_{\mu }^2-3 l_{\mu }^3+\frac{1}{2} l_{\mu }^4\right) \\ \notag
                 & \quad + C_F C_A \left(\left(\frac{691 \pi ^2}{108}-\frac{25 \pi ^4}{144}-\frac{11 \zeta_3}{12}\right)+\left(\frac{67}{6}-\frac{157 \pi ^2}{72}\right) l_{\mu }+\left(-\frac{233}{36}+\frac{\pi ^2}{6}\right) l_{\mu }^2+\frac{11}{18} l_{\mu }^3\right) \\ \notag
                 & \quad + C_F T_F \NF \left(-\frac{56 \pi ^2}{27}+\frac{\zeta_3}{3}+\left(-\frac{10}{3}+\frac{11 \pi ^2}{18}\right) l_{\mu }+\frac{19 l_{\mu }^2}{9}-\frac{2 l_{\mu }^3}{9}\right)\,,
\end{align}
where $l_\mu \coloneqq \log({s_{12}}/{\mu^2})$.

Similarly, taking the definitions from  eqs.~(50-60) from ref.~\cite{Catani:2013tia} for the $q_T$ scheme we get
  \begin{align}\begin{split}
    \delta^{(1)}_{q_T} &= C_F \pi ^2\,, \\
    \delta^{(2)}_{q_T} &= C_F^2 \left(\left(-12 \zeta_3-\frac{3}{4}+\pi ^2\right) l_{\mu }+\frac{\pi ^4}{2} \right) \\ 
                       & \quad + C_F C_A \left(\left(13 \zeta_3+\frac{245}{108}-\frac{67 \pi ^2}{24}\right) l_{\mu }-\frac{187 \zeta_3}{36}-\frac{7 \pi ^4}{48}+\frac{1181 \pi ^2}{216}+\frac{607}{81}\right) \\ 
                       & \quad + C_F T_F \NF \left(\left(\frac{5 \pi ^2}{6}-\frac{25}{27}\right) l_{\mu }+\frac{17 \zeta_3}{9}-\frac{97 \pi ^2}{54}-\frac{164}{81}\right)\,.
\end{split} \end{align}

\section{Reference Values}
\label{sec:reference-values}

To facilitate comparison with our results, we present values for
the finite remainders evaluated at a specific kinematic point. The
phase-space point we use is:
\begin{align}\begin{split}\label{eq:benchmarkphasespacepoint}
  p_1^\mu &= \{ -0.575, -0.575, 0, 0 \} \, ,  \\ 
  p_2^\mu &= \{ -0.575, 0.575, 0, 0 \} \, ,  \\
  p_3^\mu &= \{0.458858239, 0.405584802, 0.207778343, -0.053665747\} \, ,  \\
  p_4^\mu &= \{0.231129408, -0.097079562, 0.009377939, -0.209543351\} \, ,  \\
  p_5^\mu &= \{0.460012351, -0.308505239, -0.217156282, 0.263209099\} \, . 
\end{split}\end{align}
This phase-space point was chosen to match the values given in
ref.~\cite{Abreu:2020cwb} for the kinematic invariants.
For the spinors, we follow
the conventions used, for example, in ref.~\cite{DeLaurentis:2020xar},
i.e.~we take
\begin{equation}
  \label{eq:spinor_definitions}
  \lambda_\alpha = \begin{pmatrix} \sqrt{p^0+p^3}, \\ \frac{p^1+i p^2}{\sqrt{p^0+p^3}} \end{pmatrix} \, , \quad \text{and} \quad 
  \tilde\lambda_{\dot\alpha} = \begin{pmatrix} \sqrt{p^0+p^3}, \frac{p^1-i p^2}{\sqrt{p^0+p^3}} \end{pmatrix} \, .
\end{equation}
This implies that with real kinematics we have $\lambda_\alpha^*
= \tilde\lambda_{\dot\alpha}$, when the energy is positive, and
$\lambda_\alpha^* = -\tilde\lambda_{\dot\alpha}$, when the
energy is negative. 
In table \ref{tab:finite_remainders_evaluations} we provide the evaluations
of the finite remainders at the point of
eq.~\eqref{eq:benchmarkphasespacepoint}.
In addition, in \cref{tab:treeamps,tab:oneloopamps,tab:amplitudes_evaluations} 
we show the evaluations of the tree, one-loop and two-loop bare amplitudes,
which can be used to derive the results of 
\cref{tab:finite_remainders_evaluations} following the definitions in 
\cref{eq:renormalization,eq:coulourDec-gen,eq:coulourDec,eq:remainder2l}.
The one- and two-loop bare amplitudes are normalized by the following factors
\begin{equation}\label{eq:spinor-normalization-factors}
  \Phi_{+++} = \frac{[31]⟨12⟩^3⟨13⟩}{⟨14⟩^2⟨15⟩^2⟨23⟩^2} \quad \text{and} \quad
  \Phi_{-++} = A^{(0)}_{-++} = -\frac{⟨12⟩⟨23⟩^2}{⟨14⟩⟨15⟩⟨24⟩⟨25⟩} \, .
\end{equation}
The one-loop and two-loop planar remainders reproduce the values of
ref.~\cite{Abreu:2020cwb}. In this work the finite remainders were not
normalized by any spinor weight, hence a little care is needed when
comparing with table 6 of ref.~\cite{Abreu:2020cwb}.
The last four lines are the new subleading-color, non-planar
contributions.

\begin{table}[!htbp]
  \centering
  \begin{tabular}{cc}
    \toprule
    Finite Remainder & Numerical Evaluation \\
    \midrule
    $R^{(1)}_{-++}$ & $31.76842068 - 98.20723767 \, i$ \\
    $R^{(1)}_{+++}$ & $67.16227913 + 20.80380252 \, i$ \\
    \midrule
    $R^{(2,0)}_{-++}$ & $726.0944727 - 748.8429540 \, i$ \\    
    $R^{(2,0)}_{+++}$ & $1085.896384 + 310.5673949 \, i$ \\
    $R^{(2,\NF)}_{-++}$ & $-198.2921242 + 257.4649652 \, i$ \\
    $R^{(2,\NF)}_{+++}$ & $-233.4239202 - 112.3516487 \, i$ \\
    \midrule
    $R^{(2,1)}_{-++}$ & $-548.9464331 - 12.71176526 \, i$ \\    
    $R^{(2,1)}_{+++}$ & $-197.0006137 - 822.9175634 \, i$ \\
    $R^{(2,\NFqq)}_{-++}$ & $35.98395348 - 188.6772927 \, i$ \\
    $R^{(2,\NFqq)}_{+++}$ & $321.0180068 + 179.1161201 \, i$ \\
    \bottomrule
  \end{tabular}
  \caption{\label{tab:finite_remainders_evaluations}Finite remainder
    evaluations at the point of
    eq.~\eqref{eq:benchmarkphasespacepoint} with the spinors defined
    as in eq.~\eqref{eq:spinor_definitions}. The remainders are evaluated in the 
	Catani scheme as defined in the text.}
\end{table}

\begin{table}[!htbp]
  \centering
  \small
  \begin{tabular}{cc}
    \toprule
    & $\epsilon^0$\\
    \midrule
    $A^{(0)}_{-++}$ & $6.73642828 + 10.02454999 \, i$ \\
    $A^{(0)}_{+++}$ & 0 \\
    \bottomrule
  \end{tabular}
  \caption{\label{tab:treeamps}Tree amplitude evaluated at the point of
    eq.~\eqref{eq:benchmarkphasespacepoint} with the spinors defined
    as in eq.~\eqref{eq:spinor_definitions}.}
\end{table}

\begin{table}[!htbp]
  \centering
  \small
  \begin{tabular}{cccccc}
    \toprule & $\epsilon^{-2}$ & $\epsilon^{-1}$ & $\epsilon^{0}$ &
    $\epsilon^{1}$ & $\epsilon^2$ \\
    \midrule
    $A^{(1)}_{-++} \big/ \Phi_{-++}$&
    \makecell{$-1$ \\ $ $} & 
    \makecell{$-3.17428470$ \\ $- 3.14159265 \, i$} & 
    \makecell{$-3.43768120$ \\ $- 16.69077767 \, i$} & 
    \makecell{$-4.54236420$ \\ $- 48.29215997 \, i$} & 
    \makecell{$-28.34154957$ \\ $- 104.73071157 \, i$} \\[4mm]
    $A^{(1)}_{+++}  \big/ \Phi_{+++}$ &
    0 & 
    0 & 
    \makecell{$-122.48761401$ \\ $- 218.20999082 \, i$} & 
    \makecell{$-613.16200128$ \\ $- 1772.24966259 \, i$} & 
    \makecell{$-1264.78147357$ \\ $- 6727.58375625 \, i$} \\
    \bottomrule
  \end{tabular}
  \caption{\label{tab:oneloopamps}Bare, normalized one-loop amplitudes
    evaluated at the point of eq.~\eqref{eq:benchmarkphasespacepoint}.}
\end{table}

\begin{table}[!htbp]
  \centering
  \small
  \begin{tabular}{cccccc}
    \toprule
    & $\epsilon^{-4}$ & $\epsilon^{-3}$ & $\epsilon^{-2}$ & $\epsilon^{-1}$ & $\epsilon^0$\\
    \midrule
    $A^{(2,0)}_{-++} \big/ \Phi_{-++}$ &
    \makecell{$0.5$} & 
    \makecell{$2.25761803$ \\ $+ 3.14159265 \, i$} & 
    \makecell{$-3.31724534$ \\ $+ 20.90350062 \, i$} & 
    \makecell{$-55.54942677$ \\ $+ 44.34772278 \, i$} & 
    \makecell{$-248.76993460$ \\ $- 87.79211642 \, i$} \\[4mm]
    $A^{(2,0)}_{+++} \big/ \Phi_{+++}$ &
    \makecell{$0$} & 
    \makecell{$0$} & 
    \makecell{$122.48761401$ \\ $+ 218.20999082 \, i$} & 
    \makecell{$-132.67559542$ \\ $+ 2049.61318591 \, i$} & 
    \makecell{$-9927.84571218$ \\ $+ 3575.60761772 \, i$} \\[4mm]
    $A^{(2,\NF)}_{-++} \big/ \Phi_{-++}$ &
    \makecell{$0$} & 
    \makecell{$0.16666667$} & 
    \makecell{$1.33587268$ \\ $+ 1.04719755 \, i$} & 
    \makecell{$4.64626451$ \\ $+ 12.87251436 \, i$} & 
    \makecell{$10.33373683$ \\ $+ 83.15472522 \, i$} \\[4mm]
    $A^{(2,\NF)}_{+++} \big/ \Phi_{+++}$ &
    \makecell{$0$} & 
    \makecell{$0$} & 
    \makecell{$0$} & 
    \makecell{$81.65840934$ \\ $+ 145.47332721 \, i$} & 
    \makecell{$895.94750003$ \\ $+ 2327.53809534 \, i$}\\
    \midrule
    $A^{(2,1)}_{-++} \big/ \Phi_{-++}$ &
    \makecell{$0.5$} & 
    \makecell{$3.17428469$ \\ $+ 3.14159265 \, i$} & 
    \makecell{$3.54092067$ \\ $+ 26.66308714 \, i$} & 
    \makecell{$-38.30735107$ \\ $+ 112.07323410 \, i$} & 
    \makecell{$-265.12342759$ \\ $+ 331.32292317 \, i$} \\[4mm]    
    $A^{(2,1)}_{+++} \big/ \Phi_{+++}$ &
    \makecell{$0$} & 
    \makecell{$0$} & 
    \makecell{$122.48761401$ \\ $+ 218.20999082 \, i$} & 
    \makecell{$316.44565596$ \\ $+ 2849.71648558 \, i$} & 
    \makecell{$-6265.18093537$ \\ $+ 17706.12335722 \, i$} \\[4mm]
    $A^{(2,\NFqq)}_{-++} \big/ \Phi_{-++}$ &
    \makecell{$0$} & 
    \makecell{$0$} & 
    \makecell{$0$} & 
    \makecell{$0$} & 
    \makecell{$-11.30451464$ \\ $- 11.18613860 \, i$} \\[4mm]
    $A^{(2,\NFqq)}_{+++} \big/ \Phi_{+++}$ &
    \makecell{$0$} & 
    \makecell{$0$} & 
    \makecell{$0$} & 
    \makecell{$0$} & 
    \makecell{$-390.31606513$ \\ $- 1248.74833374 \, i$} \\
    \bottomrule
  \end{tabular}
  \caption{\label{tab:amplitudes_evaluations}Bare, normalized
    two-loop amplitudes evaluated at the point of
    eq.~\eqref{eq:benchmarkphasespacepoint}.}
\end{table}

\end{appendix}

\FloatBarrier

\bibliography{main.bib}

\nolinenumbers

\end{document}